\documentclass[fleqn,usenatbib]{mnras}

\usepackage{newtxtext,newtxmath}

\usepackage[T1]{fontenc}

\DeclareRobustCommand{\VAN}[3]{#2}
\let\VANthebibliography\thebibliography
\def\thebibliography{\DeclareRobustCommand{\VAN}[3]{##3}\VANthebibliography}

\usepackage{graphicx}	
\usepackage{amsmath}	
\usepackage{caption}
\usepackage{subcaption}

\newcommand{\grbfive}{{GRB\,050505}}	
\newcommand{\zabs}{{$Z_{\rm abs}$}}	
\newcommand{\zem}{{$Z_{\rm SF}\rm(SL)$}}	
\newcommand{\zte}{{$Z_{\rm SF}\rm(T_e)$}}	
\newcommand{\zsf}{{Z$_{\rm SF}$}}
\newcommand{\Te}{{$\rm{T_e}$}}	
\newcommand{\NII}{{[\ion{N}{ii}]$\lambda\lambda$6549,6584}}	
\newcommand{\NIItwo}{{[\ion{N}{ii}]$\lambda$6584}}	
\newcommand{\OII}{{[\ion{O}{ii}]$\lambda\lambda$3726,3729}}	
\newcommand{\SII}{{[\ion{S}{ii}]$\lambda\lambda$6717,6731}}	
\newcommand{\OIII}{{[\ion{O}{iii}]$\lambda\lambda$4959,5007}}	
\newcommand{\OIIItwo}{{[\ion{O}{iii}]$\lambda$5007}}	
\newcommand{\OIIIA}{{[\ion{O}{iii}]$\lambda$4363}}	
\newcommand{\OIIA}{{[\ion{O}{ii}]$\lambda\lambda$7320,7330}}	

\newcommand{\pat}[1]{{\textcolor{orange}}}

\title[Joint absorption and T$_e$-based metallicity]{First joint absorption and T$_e$-based metallicity measured in a GRB host galaxy at $z=4.28$ using \textit{JWST}/NIRSpec}

\author[A. Inkenhaag et al.]{Anne Inkenhaag$^{1}$\thanks{E-mail: ai707@bath.ac.uk},
Patricia Schady$^{1}$,
Phil Wiseman$^{2}$,
Robert M. Yates$^{3}$,
Maryam Arabsalmani$^{4,5}$,
\newauthor
Lise Christensen$^{6,7}$,
Valerio D'Elia$^{8}$,
Massimiliano De Pasquale$^{9}$,
Rub\'en  Garc\'ia-Benito$^{10}$,
\newauthor
Dieter H. Hartmann$^{11}$,
P\'all Jakobsson$^{12}$,
Tanmoy Laskar$^{13, 14}$,
Andrew J. Levan$^{14,15}$,
Giovanna Pugliese$^{16}$,
\newauthor
Andrea Rossi$^{17}$,
Ruben Salvaterra$^{18}$,
Sandra Savaglio$^{19,17,20}$,
Boris Sbarufatti$^{21}$,
Rhaana L. C. Starling$^{22}$,
\newauthor
Nial Tanvir$^{22}$,
Berk Top\c{c}u$^{1}$,
Susanna D. Vergani$^{23}$,
Klaas Wiersema$^{3}$
\\
$^{1}$ Department of Physics, University of Bath, Claverton Down, Bath BA2 7AY, UK\\
$^{2}$ School of Physics and Astronomy, University of Southampton, Southampton, SO17 1BJ, UK\\
$^{3}$ Centre for Astrophysics Research, University of Hertfordshire, Hatfield, AL10 9AB, UK \\
$^{4}$ Excellence Cluster ORIGINS, Boltzmannstraße 2, 85748 Garching, Germany \\
$^{5}$ Ludwig-Maximilians-Universität, Schellingstraße 4, 80799 München, Germany \\
$^{6}$ Niels Bohr Institute, University of Copenhagen, Jagtvej 128, DK-2200 N, Copenhagen, Denmark \\
$^{7}$ Cosmic Dawn Center (DAWN), Denmark\\
$^{8}$ Space Science Data Center (SSDC) - Agenzia Spaziale Italiana (ASI), I-00133 Roma, Italy \\
$^{9}$ Department of Mathematics, Informatics, Physics and Earth Sciences, University of Messina, Viale F. S. D’Alcontres 31, 98166 Messina, Italy \\
$^{10}$ Instituto de Astrofísica de Andaluc\'ia - CSIC, Glorieta de la Astronom\'ia s.n., 18008 Granada, Spain \\
$^{11}$ Department of Physics \& Astronomy, Clemson University, Kinard Lab of Physics, Delta Epsilon Ct, Clemson, SC 29634, South Carolina, USA \\
$^{12}$ Centre for Astrophysics and Cosmology, Science Institute, University of Iceland, Dunhagi 5, 107 Reykjav\'ik, Iceland \\
$^{13}$ Department of Physics \& Astronomy, University of Utah, Salt Lake City, UT 84112, USA \\
$^{14}$ Department of Astrophysics/IMAPP, Radboud University Nijmegen, P.O.~Box 9010, 6500 GL Nijmegen, The Netherlands \\
$^{15}$ Department of Physics, University of Warwick, Gibbet Hill Road, Coventry, CV4 7AL, UK \\
$^{16}$ Anton Pannekoek Institute for Astronomy, University of Amsterdam, P.O. Box 94249, 1090GE Amsterdam, The Netherlands \\
$^{17}$ INAF – Osservatorio di Astrofisica e Scienza dello Spazio, Via Piero Gobetti 93/3, 40129 Bologna, Italy \\
$^{18}$ INAF—Istituto di Astrofisica Spaziale e Fisica Cosmica di Milano, Via A. Corti 12, 20133 Milano, Italy \\
$^{19}$ Physics Department, University of Calabria, I-87036 Rende, Italy\\
$^{20}$ Laboratori Nazionali di Frascati, INFN (Istituto Nazionale di Fisica Nucleare), Frascati, Italy \\
$^{21}$ INAF – Osservatorio Astronomico di Brera, Via Bianchi 46, I-23807 Merate, (Lecco), Italy \\
$^{22}$ School of Physics and Astronomy, University of Leicester, University Road, Leicester, LE1 7RH, UK \\
$^{23}$ LUX, Observatoire de Paris, Université PSL, CNRS, Sorbonne Université, 92190 Meudon, France \\
}
\date{Accepted XXX. Received YYY; in original form ZZZ}

\pubyear{2025}

\begin{document}
\label{firstpage}
\pagerange{\pageref{firstpage}--\pageref{lastpage}}
\maketitle

\begin{abstract}
We present the first gamma-ray burst (GRB) host galaxy with a measured absorption line and electron temperature (T$_e$) based metallicity, using the temperature sensitive \OIIIA\ auroral line detected in the \textit{JWST}/NIRSpec spectrum of the host of GRB 050505 at redshift $z=4.28$. We find that the metallicity of the cold interstellar gas, derived from the absorption lines in the GRB afterglow, of 12 + log(O/H)$\sim 7.7$ is in reasonable agreement with the temperature-based emission line metallicity in the warm gas of the GRB host galaxy, which has values of 12 + log(O/H) = 7.80$\pm$0.19 and 7.96$\pm$0.21 for two common indicators. When using strong emission line diagnostics appropriate for high-z galaxies and sensitive to ionisation parameter, we find good agreement between the strong emission line metallicity and the other two methods. Our results imply that, for the host of GRB050505, mixing between the warm and the cold ISM along the line of sight to the GRB is efficient, and that GRB afterglow absorption lines can be a reliable tracer of the metallicity of the galaxy. If confirmed with a large sample, this suggest that metallicities determined via GRB afterglow spectroscopy can be used to trace cosmic chemical evolution to the earliest cosmic epochs and in galaxies far too faint for emission line spectroscopy, even for \textit{JWST}.
\end{abstract}




\section{Introduction}

The cosmic history of chemical enrichment is a key aspect of galaxy and stellar evolution. Obtaining accurate gas-phase metallicity measurements is necessary to trace the process of nucleosynthesis and enrichment of the interstellar medium through stellar feedback, which fuels and enriches successive generations of stars on a variety of timescales (see \citealt{Peroux2020} for a review on the baryon cycle). 

With the launch of the \textit{James Web Space Telescope} (\textit{JWST}), it has now become possible to measure the metallicity of galaxies out to higher redshift than was previously possible, which have their rest-frame optical light shifted into the thermal infrared and therefore effectively inaccessible using ground-based telescopes. Even with the \textit{Hubble Space Telescope} (\textit{HST}) the important [\ion{O}{iii}] nebular line doublet becomes inaccessible at $z\gtrsim 2.4$, and infrared spectroscopy options on {\em HST} were far more limited than on {\em JWST}. Since its launch, spectroscopic observations of large numbers of galaxies at $z>3$ have been taken with \textit{JWST}, providing metallicity measurements out to $z>8$ \citep[e.g.,][]{Arellano2022, Schaerer2022, Nakajima2023, Rhoads2023, Trump2023, Curti2023, Heintz2023, SST24, SCC25}. 

Two main methods exist for measuring the metallicity in galaxies, which each probe different components of the galaxy. One method entails using the absorption lines in the spectrum of a background quasi stellar object (QSO) or long gamma-ray burst (GRB) in the galaxy. For those sources where Ly-$\alpha$ absorption is redshifted into view for ground-based, optical telescopes ($z\gtrsim 2$) we can use absorption from singly ionised metal species to measure the metallicity of the neutral gas, \zabs. Due to their high hydrogen column densities, the clouds are self-shielded towards the high-ionisation radiation, making the low-ionisation metal absorption lines the dominant species, allowing us to measure accurate metallicities in absorption \cite[e.g.][]{Pettini1999,Prochaska2003a, Prochaska2003b, Wolfe2005,Savaglio2006,Prochaska2007a, Prochaska2007b,Fynbo2010,Fynbo2011,Rafelski2012,Fynbo2013,Krogager2013,Neeleman2013,Cucchiara2015,Bolmer2019,Heintz2023}.Because this method is not limited by the luminosity of the galaxy, it can measure metallicities even for galaxies too faint to be detected by {\em HST} and potentially even with {\em JWST} \citep{Starling05,Tanvir12,Schulze15}. This is potentially critical in understanding enrichment in more ``typical" galaxies at high redshift, and metallicities out to $z>6$ have been measured this way \citep[e.g.,][]{Kawai2006, Thone2013, Hartoog2015, Saccardi2023}. Another, and the most common, method makes use of emission lines in the spectra of star forming galaxies. These emission lines are predominantly produced by the ionised gas within the bright star-forming (SF) regions, and emission line metallicities, \zsf, are thus star formation rate weighted. This method is flux limited, as only the brightest sources will have emission lines with sufficiently high enough signal-to-noise (S/N) ratio to employ this method.

The most accurate method to measure the gas phase metallicities from emission lines is by using the temperature sensitive (\Te) auroral-to-nebular line ratios of the same ionic species, from which a reliable metallicity can be obtained, \zte\ \citep[e.g.,][]{Peimbert1967, Osterbrock1989}. However, these auroral lines are weak, the strongest one, \OIIIA\, is still $\sim 10^{-2}$ times fainter than H$\beta$. Alternative methods have thus been developed, which either use empirical or theoretical relations between various combinations of strong line (SL) ratios and \Te-based metallicities \cite[see][for a detailed review]{Maiolino2019}.

Comparing metallicities obtained through strong line diagnostics shows that \zem\ and \zabs\ (either using QSO or GRB sight-lines) of the same galaxy often do not agree, irrespective of the redshift range (compare e.g., \citealt{DeCia2018} to \citealt{Sanders2021} and \citealt{Nakajima2023}, although see 
\citealt{Christensen2004,Friis2015,Rhodin18} and \citealt{Schady2024} for examples of finding (some) agreement between \zabs\ and \zem). It is unclear whether the general disagreement has a physical origin, or if it is caused by selection effects in the galaxy sample  \cite[such as emission line spectroscopy being flux limited, and high S/N ratio absorption lines generally requiring dust-poor sightlines e.g][]{Schady2024}, the strong line relations having unaccounted third parameter dependencies \cite[e.g.][]{Kewley2002}, or \zem and \zabs\ not actually measuring the same component of the galaxy \cite[e.g.][]{Metha2020, Arabsalmani2023}. The way to solve this issue would be to measure the \Te-based metallicity for the same galaxies with available \zabs\ metallicity measurements. Indeed, for QSO-based metallicities, the lines of sight passing through the galaxies are likely illuminating different gas than the star forming gas responsible for emission line production. Using long GRBs has substantial advantages because the bursts themselves arise from the collapse of massive stars \citep[e.g.,][]{Woosley1993, Galama1998, Hjorth2003} and so should trace the same population of stars that are responsible for exciting emission lines. Whenever we mention GRBs in this paper, we specifically mean long GRBs that arise from the collapse of a massive star.

We present new observations of the host galaxy of \grbfive\ taken with \textit{JWST}/NIRSpec, in which we detect \OIIIA\, signifying the first detection of this weak line in a GRB host galaxy at $z>0.1$. These data complete the cycle-1 GRB host sample (PI: Schady; ID 2344) first presented in \cite{Schady2024}. An absorption line metallicity from the optical afterglow spectrum of \grbfive\ was reported in \cite{Berger2006} with a value of $\mbox{[M/H]}\approx -1.2$\footnote{$\mbox{[X/Y]}=\frac{\log N(X)}{\log N(Y)}-\frac{\log N(X)_\odot}{\log N(Y)_\odot}$}, corresponding to an oxygen abundance of 12 + log(O/H) $\approx 7.5$. This direct measurement of the emission line metallicity allows us, for the first time, to bridge the currently uncertain gap between the ionised and neutral ISM in a GRB host galaxy at high redshift  \cite[e.g.][]{Metha2020, Arabsalmani2023}. The host of \grbfive\ is the only host galaxy from our JWST cycle-1 proposal with detected \OIIIA\ emission and thus \citet{Schady2024} only derive an upper limit of \Te([\ion{O}{iii}])$<35000$~K for \zte\ using the stacked spectra of the other sources. 

In Section~\ref{sec:data} we describe the observations of the host of \grbfive\ and the data reduction process. We present our analysis and results in Section~\ref{sec:results} and discuss the implications of the results in context of the general GRB host population in Section~\ref{sec:discussion}. In this section we also compare the metallicities obtained through different methods, discuss the validity and value of each and we estimate the stellar mass of the host galaxy which we place in the wider context of galaxy evolution. We draw our final conclusions in Section~\ref{sec:conclusions}

We assume a standard Lambda cold dark matter ($\Lambda$CDM) cosmological model with H$_0 = 67.8$\,km\,s$^{-1}$\,Mpc$^{-1}$, $\Omega_{\rm m}=0.308$ and $\Omega_{\rm \Lambda} = 0.692$ \citep{Planck2016}. Uncertainties are 1$\sigma$ unless specified otherwise and upper limits are at the 3$\sigma$ confidence level.

\section{Data} \label{sec:data}

\subsection{Observations and Data Reduction}

\begin{figure}
 \centering
 \includegraphics[width=0.5\textwidth]{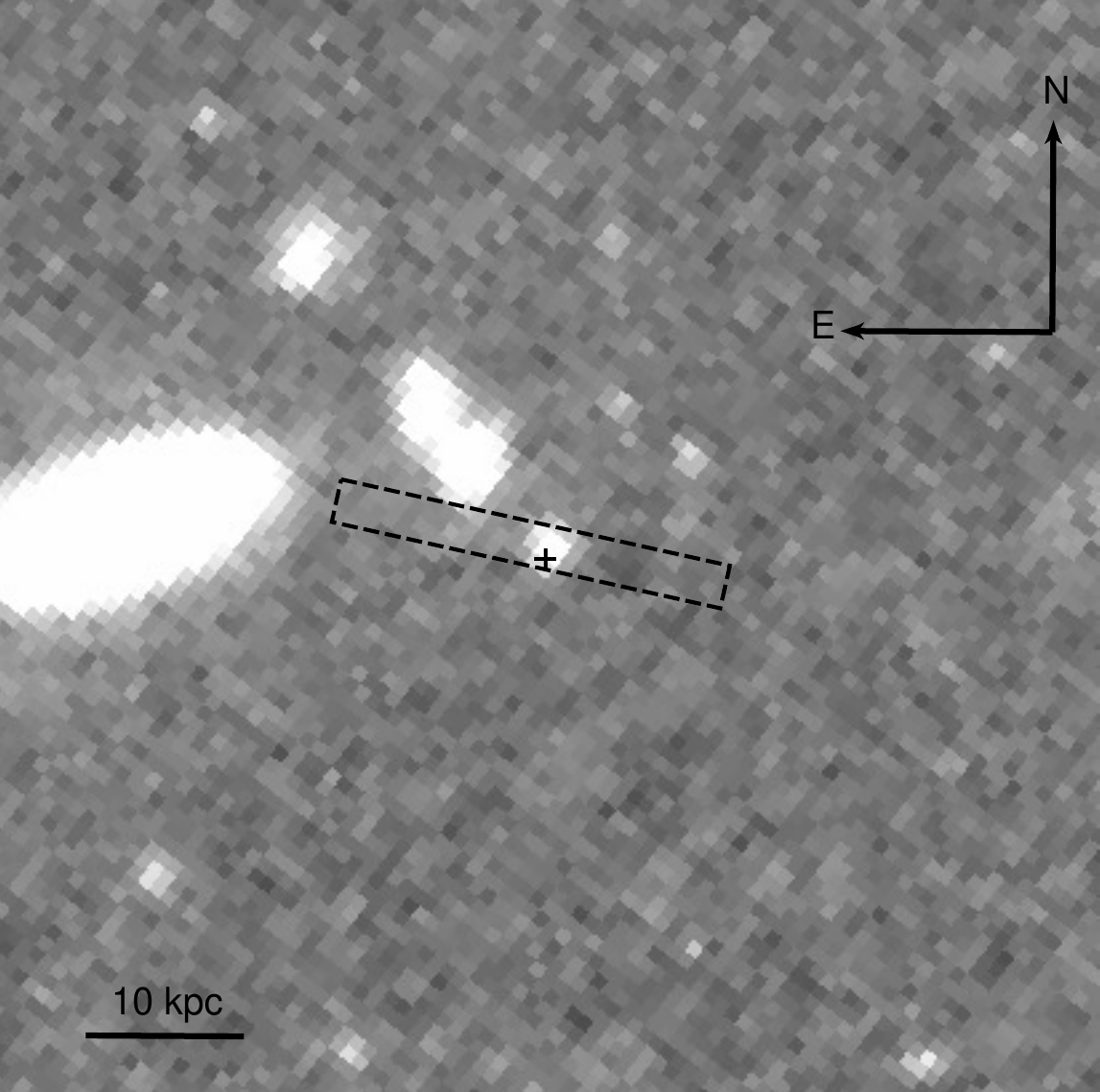}
  \caption{{\textit HST} F110W-band image of the host galaxy of GRB~050505. The position of the NIRSpec fixed slit is indicated with the dashed rectangle and the $+$ represents the GRB position. The image is oriented with north up and east to the left, and the scale of the image is given in the bottom left corner.}
 \label{fig:HSTimage}
\end{figure}

The host of \grbfive\ was observed on 2024 March 29 with the \textit{JWST}/NIRSpec S400A1 fixed slit (0\farcs4 slit width), using a two-point nod pattern. We used two grating and filter combinations (G235M/F170LP and G395M/290LP) with on source exposure times of 2042s and 584s, respectively to cover the rest-frame wavelength range between \OII\ and H$\alpha$ at the redshift of the galaxy previously determined from the GRB afterglow spectrum \citep[ $z=4.275$;][]{Berger2006}. The resolving power of the NIRSpec gratings that we used is $R=700$--$1400$, corresponding to a line width velocity dispersion of $\sigma = 90$--$200$~km~s$^{-1}$ for both grating/filter combinations. The observations were part of a larger cycle 1 \textit{JWST} campaign, and the rest of the sample, which were observed prior to the data presented here, are analysed and discussed in \cite{Schady2024}. In Fig.~\ref{fig:HSTimage} we show an {\textit HST}/F110W image (program ID 15644) centred on the host galaxy of GRB~050505 with the position of the NIRSpec fixed slit indicated with the two dashed lines. The GRB position is represented by the $+$ sign, and is accurate to within two image pixels, or $\sim 1$~kpc in physical units.

\begin{figure*}
 \centering
 \includegraphics[width=\textwidth]{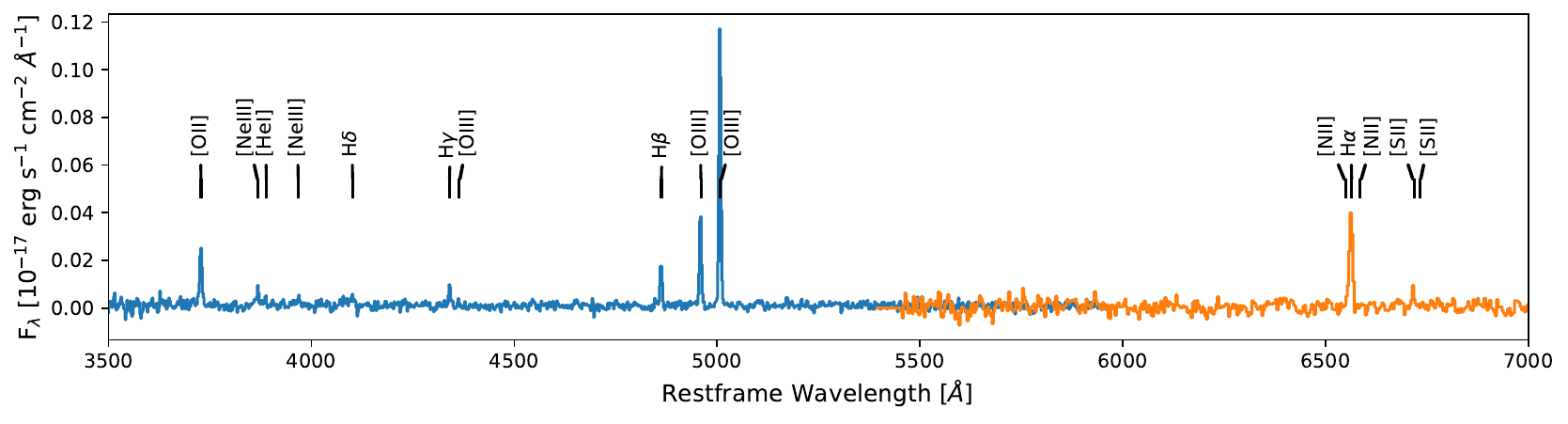}
  \caption{Spectrum of the host of \grbfive, zoomed in on the wavelength region where emission lines were detected. The blue line is the spectrum taken using the G235M/F170LP grism/filter combination and the orange line is the spectrum taken using the G395M/F290LP grism/filter combination. The spectrum has been de-redshifted to the rest frame wavelengths to allow for easier emission line identification and the emission lines listed in Table~\ref{tab:line_fits} are labeled using vertical dashes. }
 \label{fig:spectrum}
\end{figure*}

The reduced and calibrated 2D spectra were downloaded from the Mikulski Archive for Space Telescopes (MAST) Data Discovery Portal \footnote{\url{https://mast.stsci.edu/portal/Mashup/Clients/Mast/Portal.html}}. The data were reduced with version 11.17.14 of the CRDS file selection software, using context jwst\textunderscore1236.pmap, and were calibrated with version 1.13.3 of the calibration software. We extract the 1D spectrum from the 2D spectrum using the \texttt{Extract1DStep()} function from the {\sc Python JWST} pipeline \citep[v1.14.1][]{jwst_pipeline}. We show the rest-frame spectrum of the host of \grbfive\ in Fig.~\ref{fig:spectrum}, zoomed into the wavelength range where emission lines are detected.

\section{Analysis and Results} \label{sec:results}

\subsection{Line Fluxes}\label{sec:line_fluxes}

In the spectrum we detect the following lines : \OII, [\ion{Ne}{iii}]$\lambda\lambda3869, 3967$, \ion{He}{i}$\lambda3889$, H$\gamma$, \OIIIA, H$\beta$, \OIII, H$\alpha$ and [\ion{S}{ii}]$\lambda\lambda$6717, 6731. We do not detect significant emission at the position of [\ion{N}{ii}]$\lambda\lambda$6548, 6584 but we do detect a faint galaxy continuum in the bluer of the two spectra ($1.7-3.1\mu$m observer frame).

To obtain the emission line fluxes, we fit a single Gaussian to each emission line using the {\sc Python} package {\sc lmfit}. To take into account potential imperfect background subtraction and the faint galaxy continuum, we also fit a first order polynomial to create a baseline for the Gaussian fits. For greater constraint we first fit H$\beta$ and the \OIII\ doublet simultaneously, which are the lines with the highest S/N in the spectrum, and the combined fit provides additional constraint to the best-fit parameters (see Fig.~\ref{fig:spectrum}). We fit four independent parameters in this fit: the redshift of the galaxy $z$, the velocity line width $\sigma$ and two line amplitudes ($H\beta$ and [\ion{O}{iii}]~$\lambda5007$). The peak wavelengths are tied to the theoretical wavelength separation, redshifted to the observer frame. We also keep the velocity line-width ($\sigma$) tied between the lines, taking into account the NIRSpec line spread function (we assume all emission lines come from the same gas, the integrated light from the SF regions), and we fix the ratios of the amplitudes of the \OIII\ doublet to the theoretical value of [\ion{O}{iii}]~$\lambda5007/\lambda4959=2.98$ \citep{Storey2000}.  
The fits of all other lines were then fixed to the best-fit values of $\sigma = 89\pm3 $~km~s$^{-1}$ and $z = 4.27788\pm0.00003$, from the H$\beta$ and \OIII\ fit. The uncertainty on the redshift includes the NIRSpec wavelength calibration uncertainty, which is $\sim0.8$\AA \footnote{\url{https://jwst-docs.stsci.edu/jwst-calibration-status/nirspec-calibration-status/nirspec-fixed-slit-calibration-status}}. We find no strong evidence for observable differences in the Doppler shifts of the lines from different ionic species at the resolution of NIRSpec. 

\begin{figure}
 \centering
 \includegraphics[width=\columnwidth]{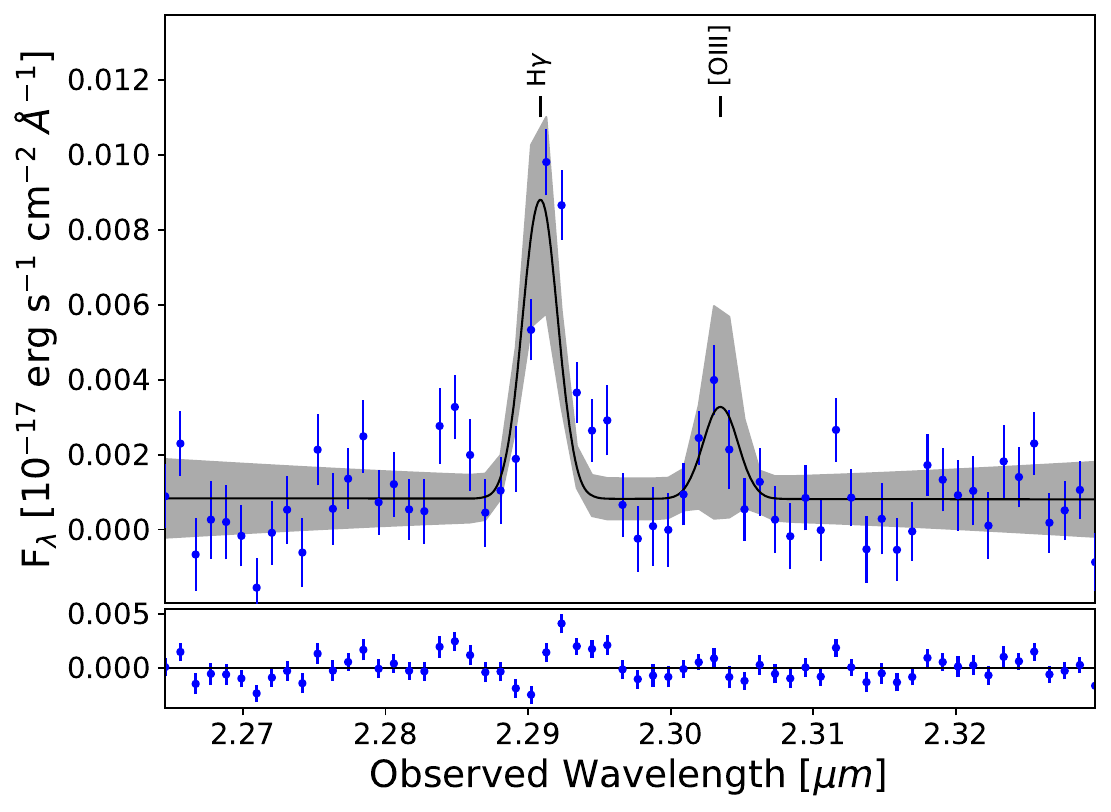}
  \caption{Fit of the H$\gamma$ and \protect\OIIIA\ emission lines of the host of \protect\grbfive. The data are shown in blue and the combined Gaussians that were fit tothe data are in black. The grey shaded area corresponds to the 3$\sigma$ error of the best combined fit to the data. The central wavelengths and line widths or all lines were fixed to the values obtained from the fit of H$\beta$ and \protect\OIII (see Fig.~\ref{apfig:line_fits}, panel 3) as these lines have the highest SNR. The centroid of both lines are marked with vertical dashes. The bottom panel shows the residuals of the data after subtracting the best fit.}
 \label{fig:OIIIA}
\end{figure}

The \OII\ doublet is blended due to the resolution of NIRSpec ($\sim23.5$~\AA\ in the G235M/F170LP grating/filter combination), which is greater than the separation of the doublet at the observed wavelength ($\sim14$~\AA\ at $z=4.28$). However, using the constraints on the redshift and the velocity width, we can fit two Gaussians to this doublet. Emission from the [\ion{Ne}{iii}]$\lambda 3967$ line in the [\ion{Ne}{iii}] doublet and the [\ion{He}{i}]$\lambda 3889$ line can also be blended with higher order Balmer lines (i.e. H$\epsilon$ and H8). We therefore tried fitting each of these lines with two Gaussians corresponding to the rest-frame peak wavelength of [\ion{He}{i}]$\lambda 3889$+H8 and [\ion{Ne}{iii}]$\lambda 3967$+H$\epsilon$. However, no significant emission was detected at the position of H$\epsilon$ or H8, and we therefore use the [\ion{Ne}{iii}] and [\ion{He}{i}] best-fit line fluxes from our single Gaussian fits to each line. For the \SII\ doublet we detect only one line above the background noise, consistent with emission from [\ion{S}{ii}]$\lambda 6717$. We therefore force the amplitude ratio between these two lines in the doublet to the maximum value of {[\ion{S}{ii}]$\lambda$6717/$\lambda$6731} = 1.4 \citep{Osterbrock2006}, resulting in a fit with reduced $\chi^2 = 1.3$ (see the bottom right panel in Figure~\ref{apfig:line_fits}). This amplitude ratio is valid for low-density environments ($n_e \lesssim 100$~cm$^{-3}$), which are common for HII regions also in GRB hosts \citep[e.g.,][]{Piranomonte2015,Izzo2017}. 
This constraint changes the flux of [\ion{S}{ii}]$\lambda$6717 from $0.30\pm0.06$~erg~s$^{-1}$~cm$^{-2}$, when fitting just a single Gaussian to this line, to $0.20\pm0.06$~erg~s$^{-1}$~cm$^{-2}$, when fitting both lines in the doublet while forcing the above amplitude ratio. 
For the \NII\ doublet we determine 3$\sigma$ upper limits on the line fluxes by forcing Gaussian fits at the expected observer-frame wavelengths, setting the line ratio to the theoretical value and fixing $\sigma$ to our previous, best-fit value. The \OIIIA\ auroral line is a weak line (see Fig.~\ref{fig:OIIIA}), as are the \SII\ lines and the \NII\ upper limit, so constraining its peak position and line width therefore improves the reliability of the flux we measure.

As a check, we also perform the fits leaving $\sigma$ as a free parameter between the fits but tying it for all sets of lines fit simultaneously. Doing this does not influence the final results significantly, so we opt for leaving it tied to the best fit from H$\beta$ and \OIII\ for consistency. We also investigate the influence of not fixing the redshift between fits, but leaving $\Delta z$ between the lines to the theoretical value. While we expect all lines to have the same, best-fit redshift, we want to investigate how much spread there would be if the redshift were left as a free parameter and how this influences the final results. Leaving $z$ as a free parameter for all lines (but keeping $\sigma$ fixed) most significantly influences the fit of the H$\gamma$ and \OIIIA\ line, with the H$\gamma$ line fitting at higher redshift and the \OIIIA\ line fitting at lower redshift than the best-fit value from the H$\beta$ and \OIII\ fit. The flux of the \OIIIA\ auroral line is increased by $\sim5$ per cent, resulting in a higher electron temperature and therefore a lower metallicity by 0.02~dex, which is still well within the uncertainty. Leaving $z$ free in the fit results in a fit with reduced $\chi^2=1.4$ compared to reduced $\chi^2=2.1$ when fixing $z$. However, since the \OIIIA\ line comes from the same atomic species as the \OIIItwo\ line, we expect it to be at the same redshift, and we therefore leave the redshift fixed to the value obtained from the \OIII\ and H$\beta$ fit. 

We correct the line flux for extinction using the {\sc python} package {\sc dust\_extinction} \citep[v1.4.1;][]{dust_ext}. We first correct the emission line fluxes for Milky Way extinction at the observed wavelength using the \texttt{G23} model \citep{Gordon2023} and setting the total-to-relative dust reddening value to $R_V=3.08$ \citep{Cardelli1989}. This dust reddening model has the advantage that it covers the whole NIRSpec wavelength range and is based on the spectroscopic extinction curves from \cite{Gordon2009, Fitzpatrick2019, Gordon2021} and \cite{Decleir2022}. It is therefore recommended by the developers when correcting for Milky Way-type extinction. We use a Galactic reddening value of E(B-V)=0.019 obtained using the {\sc python} package {\sc gdpyc} along the GRB host galaxy line of sight, with the \cite{SFD1998} dust map and corrected for the \cite{S&F2011} recalibration of this dust map. We find corrections for Milky Way (MW) extinction on the order of <1~per cent for all detected emission lines.

We then correct for the extinction in the host (at the rest fame wavelength of the emission lines) using a Small Magellanic Cloud (SMC) model. GRB host galaxies are generally low mass and metal poor and are thus more comparable to the SMC than e.g., the MW or the Large Magellanic Clouds (LMC) \citep[e.g.][]{Schady2007}. The \texttt{G24\_SMCAvg} model covers the wavelength range between $0.1-3.3$ micron. We use E(B-V)$=0.18\pm0.07$ derived from the ratio between H$_\alpha$ and H$_\beta$ and assuming a case-B Balmer decrement of 2.86 \citep{Osterbrock1989}, which is appropriate for SF regions with temperature $\sim 10^4$~K and electron densities $n_e = 10^2-10^4$~cm$^{-3}$. We find our electron temperature to be roughly consistent with this assumption (see Section~\ref{sec:results}), and while we cannot fit the electron density due to forcing the \SII\ doublet ratio to its maximum value, this ratio is valid for the lower end of the range in $n_e$ values above. The Balmer decrement given from the uncorrected fluxes for both H$\gamma$ and H$\delta$ with respect to H$\beta$ result in E(B-V) estimates $0.26\pm0.28$ and $0.01\pm0.5$, which are consistent with the value we get when using the H$\alpha$/H$\beta$ Balmer decrement. The errors are significantly larger due to the large uncertainties in the H$\gamma$ and H$\delta$ fluxes. The MW and host galaxy dust-corrected line fluxes are listed in Table~\ref{tab:line_fits}, as well as the uncorrected for both MW and host galaxy extinction line fluxes. 

\begin{table}
\caption{Results of line fitting of the emission lines of the host of \grbfive\ (both non-corrected and corrected for Milky Way and host galaxy dust) used in this work to calculate strong line ratios using $z = 4.27787\pm0.00003$ obtained from the \OIII\ and H$\beta$ fit (see the main text for details). Plots of the line fits can be found in Figures~\ref{fig:OIIIA} and \ref{apfig:line_fits} in the appendix.}
\label{tab:line_fits}
\begin{center}
\begin{tabular}{lcc}
\hline
Emission line & uncorrected flux & corrected flux\\
  & (10$^{-17}$ erg s$^{-1}$ cm$^{-2}$) & (10$^{-17}$ erg s$^{-1}$ cm$^{-2}$) \\
\hline 
[\ion{O}{ii}]$\lambda$3727 & 0.36$\pm$0.04 & 0.76$\pm$0.23\\
{[\ion{O}{ii}]}$\lambda$3730 & 0.49$\pm$0.04 & 1.05$\pm$0.30\\
{[\ion{Ne}{iii}]}$\lambda$3869 & 0.22$\pm$0.03 & 0.46$\pm$0.14\\
{[\ion{He}{i}]$\lambda$3889} & 0.11$\pm$0.03 & 0.22$\pm$0.08\\
{[\ion{Ne}{iii]}$\lambda$3967} & 0.08$\pm$0.25 & 0.16$\pm$0.07\\
H$\delta$ & 0.13$\pm$0.04 & 0.28$\pm$0.10 \\
H$\gamma$ & 0.24$\pm$0.03 & 0.49$\pm$0.14\\
{[\ion{O}{iii}]$\lambda$4364} & 0.08$\pm$0.03 & 0.15$\pm$0.07\\
H$\beta$ & 0.59$\pm$0.03 & 1.08$\pm$0.20\\
{[\ion{O}{iii}]$\lambda$4960} & 1.23$\pm$0.02 & 2.23$\pm$0.48\\
{[\ion{O}{iii}]$\lambda$5008} & 3.69$\pm$0.05 & 6.63$\pm$1.42\\
H$\alpha$ & 2.02$\pm$0.09 & 3.09$\pm$0.57\\
{[\ion{N}{ii}]$\lambda$6550} & <0.10 & <0.15\\
{[\ion{N}{ii}]$\lambda$6585} & <0.29 & <0.45\\
{[\ion{S}{ii}]$\lambda$6718} & 0.20$\pm$0.06 & 0.30$\pm$0.10\\
{[\ion{S}{ii}]$\lambda$6733} & 0.14$\pm$0.04 & 0.22$\pm$0.07\\
\hline 
\end{tabular}
\end{center}
\end{table}

\subsection{Metallicities}

\subsubsection{\Te-based diagnostics}

The measured line fluxes in galaxy spectra predominantly depend on the gas-phase abundances and temperature \citep{Peimbert1967}. Therefore, if we can measure the temperature of the gas, we can measure an accurate metallicity. More metal-poor gas is generally hotter, while metal-rich gas is generally cooler. As the \OIIIA\ auroral line is extremely sensitive to temperature, hotter gas results in a brighter \OIIIA\ line. The nebular \OIII\ lines are less sensitive to temperature and will therefore be influenced less by changes in temperature of the gas. The ratio between the nebular \OIIItwo\ and the auroral \OIIIA\ line is therefore smaller in hotter (or more metal-poor) gas. 

To calculate the electron temperature, we assume a two-zone ionisation model, in which the inner region is hotter and is traced by [\ion{O}{iii}] emission and the outer region is colder and thus traced by [\ion{O}{ii}] emission. Ideally, we would want to detect the \OIIA\ and \OIIIA\ auroral lines in order to measure \Te([\ion{O}{ii}]) and \Te([\ion{O}{iii}]) directly. These two temperatures can then be used to calculate $12 + \mathrm{log\ O^+/H^+}$ and $12 + \mathrm{log\ O^{++}/H^+}$, which combined give the average $T_e$-based metallicity, \zte$ = 12 + \mathrm{log(O/H)} = 12 + \mathrm{log(O^+/H^+ + O^{++}/H^+)}$ (we assume the contribution of O$^{+++}$ and higher ionisation states is negligible, e.g., \citealt[]{Izotov2006,Stasinska2012}). In the case where the \OIIA\ auroral lines are not detected (as in this work), we have to assume a relation between \Te([\ion{O}{ii}]) and \Te([\ion{O}{iii}]) to obtain \Te([\ion{O}{ii}]). 

Based on the nebular lines and the [\ion{O}{iii}] auroral line listed in Table~\ref{tab:line_fits}, we calculate $T_e$([\ion{O}{iii}]) and hence $T_e$([\ion{O}{ii}]) using two different methods which roughly bracket the range of values obtained from the broad literature on $T_e$([\ion{O}{iii}]) -- $T_e$([\ion{O}{ii}]) relations \citep[see e.g.,][]{Yates2020}. The first method is from \citet{Izotov2006}, using their eqns.~1 \& 2 to calculate $T_e$([\ion{O}{iii}])$ =16000\pm3000$~K and then eqn.~14 for intermediate-metallicity systems to infer $T_e$([\ion{O}{ii}])$ =14400\pm1200$~K, resulting in $12 + \mathrm{log(O/H)}=7.80\pm0.14$. Their $T_e$([\ion{O}{iii}]) -- $T_e$([\ion{O}{ii}]) relation is a second-order polynomial fit to data from the photoionization models of \citet{Stasinska&Izotov2003}. The second method is from \citet{Yates2020}, using their eqn.~3 (taken from \citealt{Nicholls2013}) to calculate $T_e$([\ion{O}{iii}])$ =16000\pm4000$ and then eqns.~9 \& 10 to iteratively obtain both $T_e$([\ion{O}{ii}])$=10000\pm3000$~K and $12 + \mathrm{log(O/H)}=7.96\pm0.21$. Their $T_e$([\ion{O}{iii}]) -- $T_e$([\ion{O}{ii}]) relation is calibrated on an observational dataset containing both individual \ion{H}{ii} region spectra and whole-galaxy spectra, and has the additional flexibility of allowing a range of $T_e$([\ion{O}{ii}]) values at fixed $T_e$([\ion{O}{iii}]) by incorporating metallicity into the fit. Eqs.~2 \& 5 from \cite{Nicholls2014} were then used in both cases to calulate \zte\ from the electron temperatures. The values obtained from these two methods are shown in Table~\ref{tab:galaxy_params}. The \citet{Yates2020} method returns an $T_e$([\ion{O}{ii}]) estimate of 10260 K (which is $\sim{}4000$ K lower than the \citealt{Izotov2006} method) and hence a higher overall $T_e$-based metallicity of 7.96 (by $\sim0.16$ dex). This difference is within the scatter in SDSS data around the \citet{Izotov2006} relation at high $T_e$([\ion{O}{iii}]) (see their fig. 4a).

\begin{table}
\caption{Electron temperatures and oxygen abundances for the host of \grbfive, using the \citet{Izotov2006} and \citet{Yates2020} methods (see main text). Upper limits on the log(N/O) ratio are also estimated using four analytic prescriptions from the literature.}
\label{tab:galaxy_params}
\begin{center}
\begin{tabular}{lcc}
\hline
Parameter & I06 & Y20 \\
\hline 
T$_e$([\ion{O}{iii}]) & $16000\pm3000$ K & $16000\pm4000$ K \\
T$_e$([\ion{O}{ii}]) & $14400\pm1200$ K & $10000\pm3000$ K \\
$12 + \mathrm{log(O^{++}/H^+)}$ & $7.69\pm0.20$ & $7.72\pm0.25$ \\
$12 + \mathrm{log(O^{+}/H^+)}$ & $7.18\pm0.16$ & $7.59\pm0.44$ \\
$12 + \mathrm{log(O/H)}$ & $7.80\pm0.14$ & $7.96\pm0.21$ \\
\hline 
log(N/O) &\multicolumn{2}{c}{$<-0.77^{(a)}$} \\
 & \multicolumn{2}{c}{$<-0.66^{(b)}$} \\
 & \multicolumn{2}{c}{$<-0.90^{(c)}$}  \\
 & \multicolumn{2}{c}{$<-0.97^{(d)}$}  \\
\hline
\end{tabular}
\end{center}
\textit{Note.} References: (a) \citet{Thurston1996} , (b) \citet{Izotov2006}, (c) \citet{Pilyugin2010}, (d) \citet{Pilyugin2016}
\end{table}

\subsubsection{Strong line diagnostics}\label{sec:SLdiags}

As in \cite{Schady2024} we use the $\hat{R}$ diagnostics from \cite{LMC24} (LMC24 from now on), and the metallicity diagnostics calibrated by \cite{NOX22} (NOX22 from now on) and \cite{SST24} (SST24 from now on) to calculate the strong line emission metallicity. Additionally, we also use \cite{Strom2018} (S18 from now on) and the recently published recalibrated $\hat{R}$ diagnostic from \cite{SCC25} (SCC25 from now on). The diagnostics that we consider are suitable for use at higher redshift because they were calibrated to conditions of high-z galaxies. They make use of the following strong line ratios:
\begin{eqnarray*}
    R_{23} &=& \frac{[\ion{O}{ii}]\lambda\lambda3726,3729 + [\ion{O}{iii}]\lambda\lambda4959,5007}{H\beta} \\
    R_3 &=& \frac{[\ion{O}{iii}]\lambda5007}{H\beta} \\
    R_2 &=& \frac{[\ion{O}{ii}]\lambda\lambda3726,3729}{H\beta} \\
    O_{32} &=& \frac{[\ion{O}{iii}]\lambda5007}{[\ion{O}{ii}]\lambda\lambda3726,3729} \\
    Ne_3O_2 &=& \frac{\ion{Ne}{iii}\lambda3869}{[\ion{O}{ii}]\lambda\lambda3726,3729}\\
    S_2 &=& \frac{[\ion{S}{ii}]\lambda\lambda6717,6731}{H\alpha}\\
    N_2 &=& \frac{[\ion{N}{ii}]\lambda6584}{H\alpha} \\
    O_3N_2 &=& \frac{[\ion{O}{iii}]\lambda5007/H\beta}{[\ion{N}{ii}]\lambda6584/H\alpha}\\
    \hat{R} &=& 0.47\times \mathrm{log}R_2 + 0.88\times \mathrm{log}R_3 
\end{eqnarray*}

\begin{table}
\caption{Metallicities values obtained using various strong line relations. The associated plots can be found in Fig.~\ref{apfig:SL_fits}.  }
\label{tab:metallicities}
\begin{center}
\begin{tabular}{lcc}
\hline
Diagnostic & Line ratio value &  12 + log(O/H)\\
\hline 
\textbf{NOX22} & & \\
\hspace{3mm} $R_{23}^\ddagger$ & $9.6\pm0.5$ & $7.90\pm0.11$ \\ 
\hspace{3mm} $R_{3}^\ddagger$ & $6.13\pm0.29$ & $7.75\pm0.07$\\
\hspace{3mm} $R_{2}$ & $1.67\pm0.21$ & $8.17\pm0.05$ \\
\hspace{3mm} $O_{32}$ & $4.19\pm0.35$ & $7.07\pm0.04$, $8.05\pm0.04$ \\
\hspace{3mm} $Ne_3O_2$ & $0.25\pm0.04$ & 7.54$^*$ \\ 
\hspace{3mm} $S_2$ & $0.17\pm0.04$ & $8.34\pm0.07$ \\
\hspace{3mm} $N_2$ & $<0.15$ & $<8.48$ \\
\hspace{3mm} $O_3N_2$ & $>42$ & $<8.34$ \\
\hline
\textbf{SST24} & & \\
\hspace{3mm} $R_{23}$ & $9.6\pm0.5$ & $7.71\pm0.13$ \\
\hspace{3mm} $R_{3}$ & $6.13\pm0.29$ & $7.59\pm0.08$, $8.26\pm0.07$\\
\hspace{3mm} $R_{2}$ & $1.67\pm0.21$ & $8.15\pm0.05$  \\
\hspace{3mm} $O_{32}$ & $4.19\pm0.35$ & $8.14\pm0.04$\\
\hspace{3mm} $Ne_3O_2$ & $0.25\pm0.04$ & $8.21\pm0.06$ \\
\hline
\textbf{S18} & & \\
\hspace{3mm} $R_{23}$ & $9.6\pm0.5$ & $8.24\pm0.13^\dagger$\\ 
\hspace{3mm} $O_3N_2$ & $>42$ & < 8.41 \\
\hspace{3mm} $N_2$ & $<0.15$ & < 8.49 \\
\hline
\textbf{LMC24} & & \\
\hspace{3mm} $\hat{R}$ & $0.80\pm0.04$ & $8.12^*$ \\ 
\hline 
\textbf{SCC25} & & \\
\hspace{3mm} $\hat{R}$ & $0.80\pm0.04$ & $8.17^*$ \\ 
\end{tabular}
\end{center}
\textit{Note.} $^*$ For thess value the line ratio in question is smaller (bigger) than the minimum (maximum) value covered by the diagnostic ($Ne_3O_{2-NOX22}<-0.41$, $\hat{R}_{LMC24}>0.76$ and $\hat{R}_{SCC25}>0.75$ at 1$\sigma$). We therefore present the turnover point between the upper and lower branch as the resulting best-fitting metallicity (see Fig.~\ref{apfig:SL_fits}. \\
$^\dagger$ In these cases the measured line ratios are larger than the maximum value, but still within $1\sigma$ of the maximum values (see Fig.~\ref{apfig:SL_fits}). We therefore assign the metallicity associated with the maximum line ratio, but also include and error.\\
$^\ddagger$ For these diagnostics we found a solution with 12 + log (O/H) < 8 using the high-EW calibration in NOX22, so we present those results. For the other NOX22 lines we used the average-EW calibration.
\end{table}

We use the fluxes listed in Table~\ref{tab:line_fits} to calculate the various SL ratios considered in this paper and then convert these to a metallicity for each calibration sample mentioned above. The plots in Appendix~\ref{apsec:metallicites} show the relations between line ratio and \Te-based metallicity for each of the diagnostics that we consider, and our measured galaxy line ratios and corresponding uncertainty are plotted as horizontal shaded regions. The metallicities that we measure for each diagnostic are listed in Table~\ref{tab:metallicities}. In those cases where the measured line ratio lies $1\sigma$ above (below) the valid range provided by each of the diagnostics, we report the metallicity corresponding to the maximum (minimum) allowed line ratio without giving errors. 

For the SST24 calibrations we limit ourselves to the metallicity range where the fits to their data have an uncertainty in (O/H) better than 0.1~dex, $7.4 < 12 + \mathrm{log(O/H)} < 8.3$. We do not detect the continuum emission of the galaxy in the spectrum, so we are unable to determine the equivalent width (EW) of the emission lines, and we can therefore not use the EW(H$\beta$)-dependent diagnostics from NOX22. NOX22 argue that EW(H$\beta$) can be used as a tracer of the ionisation state of the galaxy ISM, with low-EW(H$\beta$) corresponding to low ionisation and high-EW(H$\beta$) corresponding to high ionisation. In the early universe, ionisation was higher \citep[e.g.,][]{Kewley2013a, Kewley2013b, Steidel14, Strom2018}, and thus NOX22 argue that the high-EW(H$\beta$) calibrations are more appropriate for high-$z$ galaxies. Using a sample of ten galaxies at $z = 4-8.5$, \cite{Nakajima2023} confirmed that the high-EW(H$\beta$) calibrations are indeed more accurate for high-$z$ galaxies, irrespective of the actual EW(H$\beta$) value. The ionization ratio in the host of \grbfive\ is also high (log(\OIIItwo/\OII)$\sim0.6$, {similar} to e.g., the high excitation sample from \citealt{Stasinska2015}) and we therefore use the high-EW(H$\beta$) calibrations for the host of \grbfive\ despite not being able to measure EW(H$\beta$). The high-EW calibrations are only valid for 12 + log(O/H) $\leq 8$, and thus for those NOX22 diagnostics where we measure 12 + log(O/H) > 8, we then re-calculate the metallcitity using the EW-averaged diagnostics.

For diagnostics that use the \NIItwo\ line, which are the $N_2$ and the $O_3N_2$ diagnostics, the metallicities listed are upper limits as we only have an upper limit for this emission line (see Table~\ref{tab:line_fits} and Fig.~\ref{apfig:line_fits}). The $N2$ diagnostic is linear between 12+log(O/H)$\sim7.8-8.7$ and although it is very sensitive to ionisation parameter \citep[e.g.,][]{Kewley2002}, it can still be used to discriminate between double-branched solutions. The same argument holds for the $O_3N_2$ diagnostic. For the SST24 calibration, we were unable to discriminate between the two solutions for $R_{3}$ (see Fig.~\ref{apfig:SL_fits}), and in Table~\ref{tab:metallicities} we therefore give both the lower and the upper branch metallicities. 

To determine the error on the metallicity calculations we perform Markov-chain Monte Carlo in which we randomly draw values from the observed line ratios assuming a normal distribution with standard deviation equal to our measured line ratio standard deviations and compute the metallicity using the drawn line ratio. This is repeated 5000 times after which we use the median and the standard deviation for the distribution in metallicities to determine the most likely metallicity and $1\sigma$ uncertainty given the measured line ratios and uncertainties. If the diagnostic is double branched we repeat this for both branches and report both values in Table~\ref{tab:metallicities}.

Only NOX22 lists uncertainties on their diagnostics in the metallicity direction for all line ratios used in their table~4, which vary between 0.09~dex for $R23$ (high-EW) and 0.60~dex for Ne3O2 (average for all EWs). We only use the SST24 diagnostics in the range of metallicities for which they state the median 12+log(O/H) uncertainty is <0.1 dex, so we use 0.1~dex as the systematic error on the SST24 calibrations. For the remaining calibration samples no uncertainties were given, and hence we adopt a systematic uncertainty of 0.2~dex {for each diagnostic}, which seems reasonable based on the systematic uncertanties from NOX22 and SST24. {These systematic errors represent the scatter in the calibration sample about the best-fit diagnostic.} We add the systematic uncertainty of the calibration in quadrature to obtain the full uncertainty on the metallicity. In the cases where we present the metallicity without error in Table~\ref{tab:metallicities}, we only use the systematic error on the calibration to compute the full uncertainty of the metallicity.

\section{Discussion} \label{sec:discussion}


\subsection{Host of GRB~050505 in the general GRB host galaxy sample}

\subsubsection{Electron temperatures}

We compare the electron temperature of the host of \grbfive\ to that of other GRB host galaxies, GRB980425 \citep[\Te(\lbrack\ion{O}{iii}\rbrack) = 10500 $\pm$ 0.0500~K at $z = 0.0086$;][]{Kruhler2017}, GRB031203 \citep[\Te(\lbrack\ion{O}{iii}\rbrack) = 13400 $\pm$ 0.2000~K, $z = 0.1055$;][]{Prochaska2004}, GRB060218 \citep[\Te(\lbrack\ion{O}{iii}\rbrack) = 24800$^{+5000}_{-3000}$~K, $z = 0.03342$;][]{Wiersema2007} and GRB100316 \citep[\Te(\lbrack\ion{O}{iii}\rbrack) = 11900$\pm$800~K and \Te(\lbrack\ion{O}{ii}\rbrack) = 10400$\pm$1100~K, $z = 0.0591$;][]{Starling2011}. While the value we measure for both \Te([\ion{O}{iii}]) as well as \Te([\ion{O}{ii}]) is higher than this small sample of GRB hosts, it is still generally consistent within 1$\sigma$. Nevertheless, the host galaxy of \grbfive\ is at significantly higher redshift than our small comparison sample ($z<0.11$), 
and \Te\ is generally observed to be lower in the local SF galaxy population \citep[see e.g.,][]{Shi2014, Hirschauer2015, Yates2020} than the high-z SF population \citep[see e.g.,][]{Schaerer2022,Rhoads2023}. This is due to galaxies at lower redshift having higher metallicities because of galactic chemical enrichment over time. The \Te\ values of the four local GRB hosts mentioned above are consistent with local SF galaxy samples (typically 7000--15000~K; e.g., \citealt{Shi2014, Hirschauer2015, Yates2020, Rogers2022}, with the occasional exception of higher temperatures around 23000~K such as observed in \citealt{Hirschauer2015} and for GRB~060218 above). The host of \grbfive, {on the other hand,} agrees better with the higher temperatures typically observed in high-z SF galaxies (>15000~K; e.g., \citealt{Christensen2012,Patricio2018,Schaerer2022,Rhoads2023,SST24,LMC24}), although that sample is still small (on the order of tens of galaxies). 

\subsubsection{Galaxy characteristic properties}

While the distribution of GRB host galaxy metallicities at $z\lesssim 3$ is offset towards lower metallicities when compared to the general population of SF galaxies at comparable redshifts, \citep[e.g.,][]{Graham2013,Kruler2015, Graham2017, Arabsalmani2018, Palmerio2019}, GRB hosts do appear to follow the general trend in the mass-metallicity relation of SF galaxies \citep[e.g.,][although see \citealt{Graham2023}]{Vergani2015,Arabsalmani2018}. 

There is only a single \textit{HST} WFC3/F110W observation of the host galaxy of \grbfive. Photometry of this host galaxy in a 0.4 arcsecond aperture provides an AB magnitude F110W=25.90$\pm$0.06. This corresponds to an absolute magnitude of $M_{2000\AA} = -20.3$ and a UV indicated star formation rate of $\sim 1.5$ M$_{\odot}$ yr$^{-1}$ \citep[using the normalisation of][]{Hirashita03}). we note that the host galaxy is relatively luminous compared to other GRB hosts at $z>4$ \citep{Tanvir12,Schulze15,McGuire16}.

In order to measure the galaxy stellar mass, M$_\star$, from the galaxy spectral energy distribution (SED), we first convolve the G235M/F170LP NIRSpec spectrum with the NIRCam F200W and F277W response curves. This results in measured AB magnitudes of F200W$= 25.36\pm0.20$ and F277W$=24.66\pm0.15$ mags, which we combine with the HST/WFC3 F110W photometry to produce a UV/optical SED. After correcting for the Galactic foreground extinction, we modeled the SED  using the code {\sc Cigale}\footnote{\url{https://cigale.lam.fr/}} \citep{Boquien2019} with a redshift fixed to $z=4.27$. Although the SED is too sparsely sampled to fully resolve the degeneracy between attenuation, age, and SFR, the \textit{JWST}-estimated magnitudes above the Balmer break allow us to constrain the stellar mass to $\log(M_\star/M_\odot) = 9.4 \pm 0.4$.

Using the dust corrected H$\alpha$ flux we calculate the star formation rate \citep{Kennicutt1998} of the host of \grbfive, assuming the \cite{Chabrier2003} initial mass function. We find a star formation rate (SFR) of $26\pm5$~M$_\odot$~yr$^{-1}$, which is consistent with the distribution of SFRs measured in larger samples of GRB host galaxies \citep[e.g.,][]{Christensen2004, Kruler2015,Palmerio2019, Schneider2022}. Combined with the stellar mass we find, the SFR suggests that the host of \grbfive\ lies on the main sequence for SF galaxies \citep[e.g,][]{Popesso2022, Curti2024}. The combination of SFR and \Te-based we measure are also consistent with the fundamental metallicity relation from \cite{Sanders2020}. We therefore conclude that our GRB host galaxy is consistent with the general population of SF galaxies.

The sample of $\sim 100$ GRB host galaxies from \cite{Perley2016b} show a wide range of stellar masses (see their fig.~3), but the sensitivity of their {\em Spitzer} observations did not allow them to probe galaxies at $z\gtrsim4$ with stellar masses much smaller than $<10^{10}~{\rm M}_\odot$. Similarly, there are not many GRB hosts with metallicities measured as low as 12+log(O/H)$\sim$7.8 for \zte. This is likely due to selection effects because emission line spectroscopy is only available for the brightest sources (which tend to be more massive) and the \OIIIA\ auroral line is weak, needing bright sources and longer integration times. It is only because of our sensitive \textit{JWST} spectra that we are able to measure oxygen abundances down to far lower metallicities than previously possibly for other GRB hosts at comparable redshift.

\subsection{Metallicity Relations}

\begin{figure}
 \centering
 \hspace*{-.8cm}\includegraphics[width=.52\textwidth]{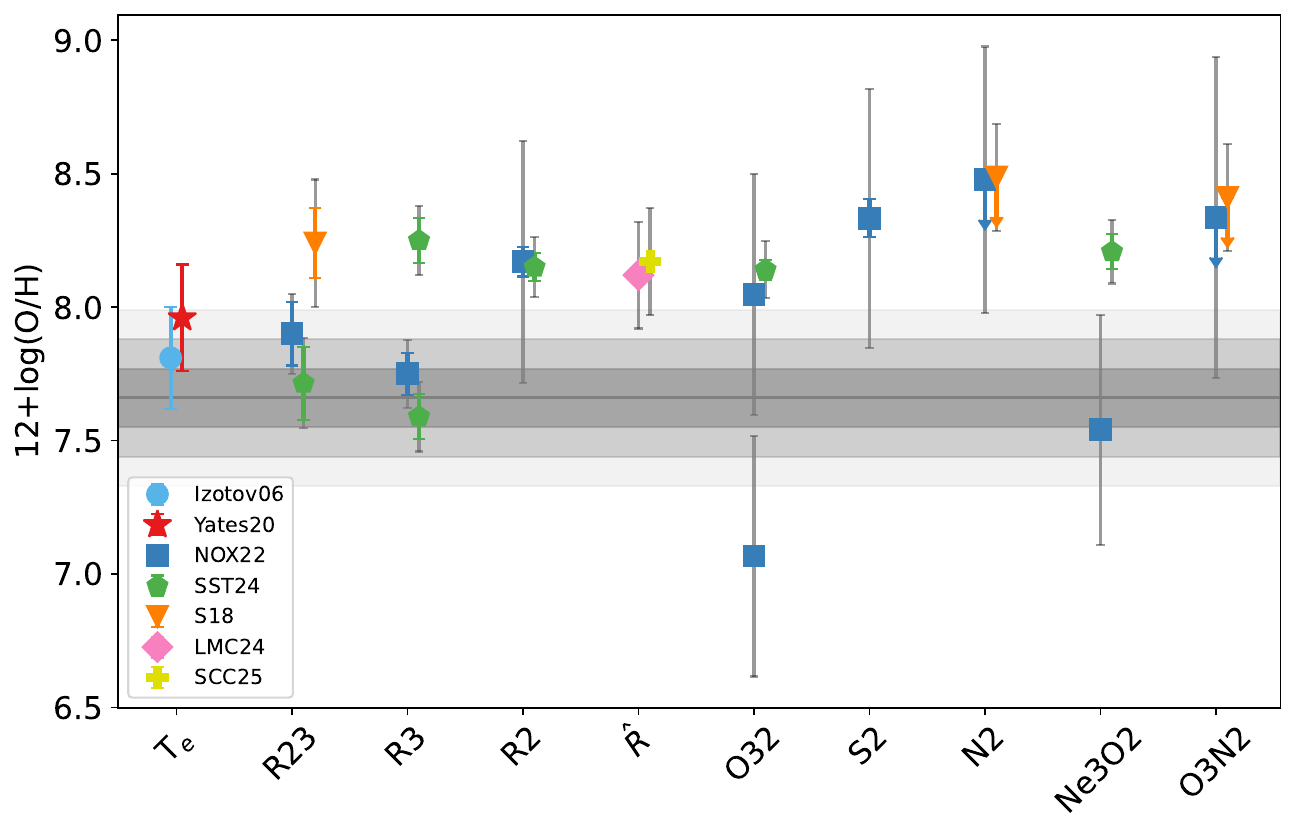}
  \caption{Comparison of the metallicities listed in Table~\ref{tab:metallicities} and the re-calculation of \zabs\ and both methods of obtaining \zte. The solid grey line corresponds to \zabs, whereas the shaded regions represent the $1\sigma$, $2\sigma$ and $3\sigma$ confidence regions in progressively lighter shades of grey. Different calibrations are represented using different colours and marker styles, with the coloured (shorter) error bar representing the statistical uncertainty and the larger dark grey error bar representing the full uncertainty including the systematic uncertainty in the SL diagnostics. The markers without coloured error bars are the values in Table~\ref{tab:metallicities} presented without the errors because the peak of the diagnostic was outside of the 1$\sigma$ error of the line ratio, the dark grey error bars represent the systematic uncertainty in the calibrations in these cases. In cases where it was not possible to discriminate between the lower or upper branch, we plot both metallicity solutions. For a given line ratio the data points corresponding to different calibrations have been slightly offset to each other for clarity.
  }
 \label{fig:z_comp}
\end{figure}

The collapsar model \citep{Woosley1993} for long GRBs predicted an association between long GRBs and massive stars, and observations confirmed this \citep[e.g.,][]{Galama1998, Hjorth2003}. This means their sight-line emanates from the same SF regions that dominate the emission line spectra. We can use the absorption imprint left on the afterglow spectra of these long GRBs to probe the neutral and low-level ionised gas metallicity in the GRB surroundings \citep[e.g.,][]{Kruhler2017, Wiseman2017a}. Typically, the closest absorbing clouds have been found to lie at a few hundred parsec from the GRB (\citealt{Vreeswijk2007, Vreeswijk2013, d'elia2014}; although see \citealt{Saccardi2023}). Using GRB absorption spectra therefore ensures we have two metallicity measurements along the same line of sight, probing the same region of the galaxy, allowing for a comparison between \zsf\ and \zabs\ in the same region. We are unsure whether the metallicity of the ISM in the neutral and ionised phase is comparable, and our \textit{JWST} data allow us to examine this. \cite{Schady2024} investigated the relation between \zabs\ and \zem, but in this work, for the first time, we can compare the metallicity of both gas phases using direct measurements. 

Fig.~\ref{fig:z_comp} shows all emission line metallicities listed in Table~\ref{tab:metallicities} and the \Te-based metallicities listed in Table~\ref{tab:galaxy_params} for comparison. \zabs\ is indicated by the solid grey line, with progressively lighter shaded regions corresponding to the $1\sigma$, $2\sigma$ and $3\sigma$ uncertainty region, respectively. When comparing the different strong line metallicities in Fig.~\ref{fig:z_comp}, we can see that there is a wide spread in metallicities we find from the different SL diagnostics, both within the same calibration sample (marked with the same colour and marker) and between different calibration samples. This is a common issue when using SL diagnostics, with variations as large as $\sim0.6$~dex between different diagnostics reported before \citep[e.g.,][]{Kewley2008, Teimoorinia2021}. The SL relations have traditionally been calibrated on local samples of galaxies which may not be representative of early populations. For example, galaxies in the early universe have younger stellar populations and higher ionisation parameters than local galaxies \citep[e.g.,][]{Steidel14}. Since the launch of \textit{JWST}, there have been efforts to calibrate these relations for ISM conditions in high-$z$ galaxies \citep[e.g.,][]{Hirschmann2023, LMC24, SST24}, but these are based on small samples and even for the high-$z$ calibrations there is a spread in metallicities between the different line ratios of up to $\sim0.5$~dex \citep[e.g.,][]{NOX22, Nakajima2023, LMC24, SST24}. We refer the reader to the literature on the differences in calibration samples and diagnostics, see e.g., \cite{Kewley2008}, \cite{Maiolino2019}, \cite{SST24} and \cite{LMC24} for extensive reviews, and refrain from commenting on what can be the cause of the differences between the obtained strong line metallicity values.

\subsubsection{Absorption vs. T$_e$-based metallicities}

\begin{figure}
 \centering
 \includegraphics[width=.5\textwidth]{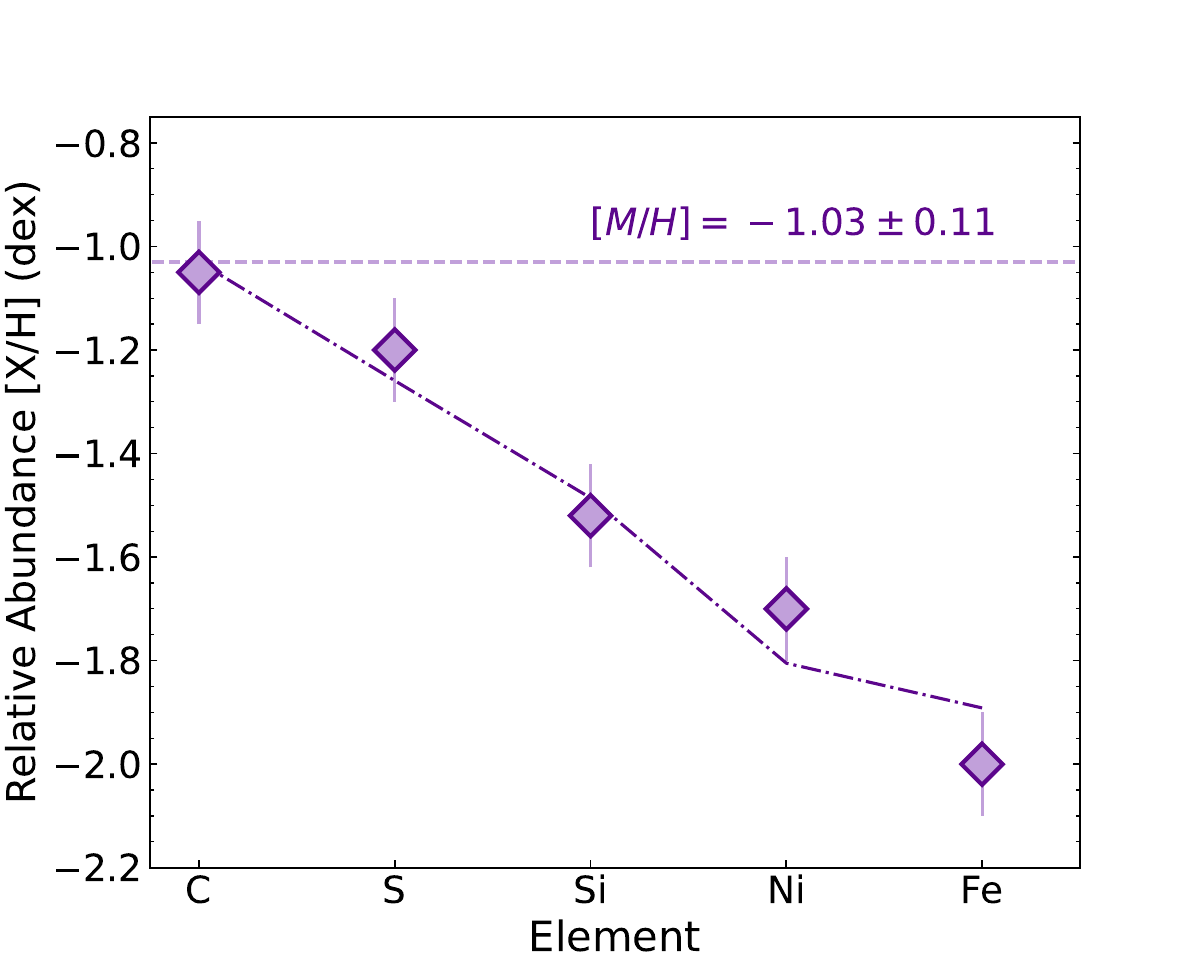}
  \caption{Dust depletion fits to the neutral element relative abundances measured in the LRIS optical afterglow spectrum of \grbfive\ \protect\citep{Berger2006}. We assume an uncertainty of 0.1~dex for all column densities. The dot-dashed line corresponds to the best-fit depletion pattern, which results in a best-fit, dust-depletion corrected absorption metallicity of $\mbox{[M/H]}=-1.03\pm 0.11.$}
 \label{fig:abs_met}
\end{figure}

The afterglow spectrum used to obtain the absorption based metallicity of the host of \grbfive\ was obtained with the Low Resolution Imaging Spectrometer (LRIS) on the Keck I 10m telescope. Since this is a low resolution instrument, the absorption lines used to calculate the metallicity can suffer from hidden saturation, and the metallicity should thus be considered a lower limit. The lower limit of 12 + log(O/H) $> 7.5$ presented in \cite{Berger2006} \cite[assuming a solar abundance 12+log(O/H)=8.69;][]{Asplund2009} is consistent with the range in $T_e$-based metallicities that we measure ($12+\log({\rm O/H})=7.80-7.96$; see Table~\ref{tab:metallicities}).

To try and constrain better the absorption based metallicity, we re-calculated the metallicity following the procedure of \cite{Wiseman2017a}. This method uses multiple, singly ionised metal absorption lines to simultaneously constrain the dust depletion and neutral gas metallicity \zabs, making use of the detailed depletion patterns from \cite{DeCia2016}. Although the measured column densities likely suffer from some level of hidden saturation, by fitting all relative abundances simultaneously, we mitigate some of the uncertainties associated with low resolution absorption line measurements. Our best dust depletion fits to the relative abundances taken from \citet{Berger2006} are shown in Fig.~\ref{fig:abs_met} and give a best fit metallicity of $\mathrm{[M/H]} = -1.03\pm0.11$ corresponding to an oxygen abundance $12 + \mathrm{log(O/H)} = 7.66\pm0.11$. \citet{Berger2006} state that \ion{S}{ii} and \ion{Ni}{ii} are likely not saturated, which is consistent with the relative abundances of these two species lying above the best fit dust depletion model shown in Fig.~\ref{fig:abs_met}. In such a case, the relative abundances that lie below the line of best fit should be considered lower limits. Nevertheless, the consistency between the model and all data points gives some validity to the absorption metallicity that we measure. 

Interestingly, our absorption line metallicity is consistent within errors with the two $T_e$-based metallicities we measure (see Fig.~\ref{fig:z_comp}). If the absorption line metallicity were considerably larger than what we measure, the implication would be that the neutral phase ISM is more enriched than the gas within star forming regions, contrary to theoretical expectations \cite[e.g.][]{Metha2020,Metha2021,Arabsalmani2023}.

As shown in Fig.~\ref{fig:z_comp}, the $R23_{SST24}$ diagnostic and the lower branch solution of $R3_{SST24}$ are also consistent with this improved value of the absorption metallicity within 1$\sigma$, as well as the $R3_{NOX22}$ diagnostic. We note that while the $Ne3O2_{NOX22}$ is consistent with \zabs, this is a data point plotted without error, because the line ratio was more than $1\sigma$ below the minimum value applicable for the diagnostic. Additionally, the relation between $Ne3O2$ and metallicity is almost flat (see Fig.~\ref{apfig:SL_fits}), and thus insensitive to changes in the line ratio. We therefore do not consider this a reliable diagnostic. $R23_{NOX22}$ is consistent at 1.2$\sigma$ and while this is not quite $1\sigma$ significance, it is still notably more consistent than the other diagnostics not mentioned already. 

\subsubsection{T$_e$-based vs. strong line metallicities}

Comparing the different SL diagnostics plotted in Fig.~\ref{fig:z_comp} to the two \Te-based metallicities, we see that the $R23$ and $R3$ diagnostics generally agree best. While most, if not all, of the SL diagnostics are technically consistent with the \Te-based metallicities due to the large systematic uncertainties in the calibrations (the grey errorbars), most diagnostics result in metallicities higher than the \Te-based metallicity. 

The $R23$ diagnostic is relatively independent of the ionisation parameter because it uses both the singly and doubly ionised lines. Although not completely independent \citep[see e.g.,][]{Kewley2002}, we do generally see it agrees better with our value of \zte\ than the diagnostics heavily dependent on the ionisation parameter, such as $N2$ and $S2$ which only use either just the singly or just the doubly ionised ions of one species. We do note that because of the spread in wavelengths in the emission lines needed for the $R23$ diagnostic, it is more sensitive to the reddening correction used compared to diagnostics that use emission lines closer together in wavelength.

The NOX22 $R23$ and $R3$ high-EW diagnostics as presented in Table~\ref{tab:metallicities} and plotted in Fig.~\ref{fig:z_comp} are consistent with \zte\ within 1$\sigma$, indicating that the high-EW relations are the best calibration for high-z galaxies presented by NOX22, as also suggested by \cite{Nakajima2023}. When using the averaged $R23$ and $R3$ relations from NOX22, the agreement is also better than the other diagnostics, but considering the EW dependence improves both diagnostics further compared to the \Te-based metallicity. Interestingly, the LMC24 $\hat{R}$ diagnostic deemed best by \cite{Schady2024} for the rest of the GRB host galaxy sample only barely agrees with \Te\ within 1$\sigma$ and does not agree with \zabs. 

\subsubsection{The N/O--O/H relation}

The N/O--O/H relation in the local universe shows a relatively flat relation at low metallicities that then starts increasing as a power law at $12+\log({\rm O/H})\sim 8.0$ with the onset of secondary N production \citep[e.g.,][]{Edmunds1978,Henry1999,Koppen2005,Vincenzo2016}. With JWST, some high redshift galaxies have been observed to have surprisingly high N/O ratios relatively to their metallicity \cite[e.g.][]{Cameron2023,MarquesChaves2024}. We thus calculate log(N/O) for the host galaxy of GRB~050505 using various methods (see Table~\ref{tab:galaxy_params}). Two of these methods (from \citealt{Thurston1996} and \citealt{Izotov2006}) explicitly depend on the electron temperature, which is calculated using observed auroral/strong emission lines and fits to photoionisation models. The other two methods (from \citealt{Pilyugin2010} and \citealt{Pilyugin2016}) depend only on oxygen and nitrogen (and sulphur, in the case of \citealt{Pilyugin2010}) strong-line ratios, with empirically derived N/O relations fit to a sample of precisely selected H\textsc{ii} regions. The various estimated upper limits range from $\textnormal{log(N/O)} < -0.97$ to $-0.66$ and are thus consistent with the general trend between 12+log(O/H) and log(N/O) seen in star-forming galaxies \citep[e.g.,][]{Dopita2016,Nicholls2017}, unlike other notable high-redshift systems such as GN-z11 \citep{Cameron2023}.

\subsection{Mixing of Metals}

The consistency between \zte\ and \zabs\ within 1$\sigma$ suggests that metals newly synthesised by stars are efficiently distributed within the star forming regions (traced by \zte) and the neutral ISM (traced by \zabs). In order to investigate the expected differences between absorption and emission line metallicities \cite{Arabsalmani2023} used the EAGLE cosmological hydrodynamical simulations to study the predicted relation between the metallicity within star forming regions and the metallicity along random sightlines as a function of galaxy and sightline properties. They found the closest agreement between the two probes for sightlines that crossed close to the galaxy centre, within $1-2$~kpc, and this agreement improved further for sightlines that probed larger column densities. However, even for high column density and low radial offset sightlines, \citet{Arabsalmani2023} predicted a small offset on the order of $\sim0.2$~dex between \zabs\ and \zsf, with \zsf\ being the higher value of the two. The biggest difference was found when the simulated sightlines were at large radial offsets from the galaxy centre, which can be understood if the outskirts of galaxies are less enriched in metals than the galaxy disk, where the majority of star formation (and thus nucleosythesis) occurs. The consistency that we measure between \zabs\ and \zsf\ for the host galaxy of \grbfive\ is in agreement with the predictions from \cite{Arabsalmani2023}, given the large hydrogen column density measured along the GRB line of sight \cite[$\log(N_H)/{\rm cm}^{-2}=22.1$;][]{Berger2006} and the small GRB positional offset from the galaxy centre (see Fig.~\ref{fig:HSTimage}).

\cite{Metha2023} (whose work builds on \citealt{Metha2020} and \citealt{Metha2021}) carried out a similar analysis, comparing results between a number of cosmological simulations (Illustris, IllustrisTNG and EAGLE), and find similar conclusions that there is an offset between \zabs\ and \zsf, for all three simulations. They also find that the difference between \zabs\ and \zsf\ increases when the metallicity decreases, in agreement with \cite{Arabsalmani2023}, although \cite{Metha2023} do not consider the effect of impact parameter or the line of sight column density.

Instead \cite{Metha2023} considered the effect of introducing a metallicity cutoff in the GRB progenitor, and they found that the difference between \zsf\ and \zabs\ increasingly deviated for galaxies with \zsf\ larger than the imposed metallicity cutoff. This can be understood since GRBs with high metallicity hosts would be more likely to occur at large radial offsets, where the gas-phase metallicity is expected to be lower due to metallicity gradients.
This therefore implies that finding \zsf>\zabs\ from observations could be evidence for the existence of a metallicity bias in GRB progenitors, although a large sample of \zsf\ and \zabs\ pairs would be required to convincingly detect such an offset.

\section{Summary and Conclusions} \label{sec:conclusions}

In this paper we present the first GRB host galaxy for which we have a metallicity measurement probing the warm and the cold ISM, for the first time allowing to bridge the gap between these two phases of the ISM. In our \textit{JWST}/NIRSPEC spectra, we detect the \OIIIA\ auroral line and use it to calculate the electron temperature of the emitting gas and hence the $T_e$-based metallicity from emission lines. This is the most direct, mostly model-independent method of determining the metallicity. Comparing it to the model dependent strong line metallicites and the non-flux limited GRB afterglow absorption based metallicity, we see the \Te-based metallicity seems to agree better with \zabs, although the SL diagnostics calibrated on ISM conditions in the early universe do also agree with the \zte. The agreement between \zabs\ and \zte\ suggest mixing between the neutral and ionised gas is efficient along the line of sight. 

Using the GRB afterglow spectrum combined with the integrated spectrum eliminates the offset between where the two metallicities are measured, and therefore this cannot be the cause of the discrepancy between \zem\ and \zabs\ in this work. This leaves a physical reason, or a systematic error in the calibrations of the SL diagnostics. We find the SL diagnostics independent from the ionisation parameter (e.g., R$23$) or diagnostics where a dependency on the ionisation parameter is explicitly included (as in NOX22) agree better with our measurements of \zte. We therefore advocate for using these specific diagnostics when determining the metallicity for high-z galaxies. However, whenever possible, we suggest using \zte\ or \zabs, since they are less dependent on models, calibration samples and additional parameters such as ionisation.

Our improved \zabs\ is in good agreement with \zte\ and \zem\ when using the $R23$ diagnostic, which could imply that \zabs\ can be used to trace metallicity of SF regions in high-z galaxies. To confirm this result, and to investigate further if the \Te-based metallicity indeed traces closest the neutral gas metallicity obtained through GRB absorption line spectra, the sample of high-$z$ GRB host galaxies with detections of the \OIIIA\ auroral emission lines has to be expanded, although significant efforts have been made over the last few years. If indeed confirmed with a large sample, our results could imply that GRB absorption spectroscopy can be used to trace cosmic chemical evolution to the earliest cosmic epochs and for galaxies too faint for emission line spectroscopy, even using \textit{JWST}.

\section*{Acknowledgements}
Support for program 2344 (PI. Schady) was provided by NASA through a grant from the Space Telescope Science Institute, which is operated by the Association of Universities for Research in Astronomy, Inc., under NASA contract NAS 5-03127.
AI and PS acknowledge support from the UK Science and Technology Facilities Council, grant reference ST/X001067/1. PW acknowledges support from the UK Science and Technology Facilities Council, grant reference ST/Z510269/1.
RGB acknowledges financial support from the Severo Ochoa grant CEX2021-001131-S funded by MCIN AEI/10.13039/501100011033 and PID2022-141755NB-I00.
RLCS acknowledges support from the Leverhulme Trust grant RPG-2023-240.
AR acknowledges support from PRIN-MIUR 2017 (grant 20179ZF5KS).
Funded by the Deutsche Forschungsgemeinschaft (DFG, German Research Foundation) under Germany´s Excellence Strategy – EXC 2094 – 390783311.
SDV acknowledges the support of the French Agence Nationale de la Recherche (ANR), under grant ANR-23-CE31-0011 (project PEGaSUS)
This work makes use of Python packages {\sc numpy} \citep[v1.26.4;][]{2020Natur.585..357H}, {\sc uncertainties} (v3.2.2; \url{http://pythonhosted.org/uncertainties/}), {\sc matplotlib} \citep[v3.9.1;][]{2007CSE.....9...90H}, {\sc gdpyc} (v.1.1; \url{https://gdpyc.readthedocs.io/en/latest/}) and {\sc lmfit} \citep[v1.3.2;][]{lmfit}.
This work made use of Astropy (v6.1.2; \url{http://www.astropy.org}): a community-developed core Python package and an ecosystem of tools and resources for astronomy \citep{astropy:2013, astropy:2018, astropy:2022}.

\section*{Data Availability}

The \textit{JWST} data is publicly available through the MAST archive. Code used for this analysis will be made available on reasonable request to the corresponding author.



\bibliographystyle{mnras}
\bibliography{references} 




\appendix


\section{Emission line fits}\label{apsec:linefits}

\begin{figure*}
 \centering
 \hspace*{-.5cm}\includegraphics[width=0.5\textwidth]{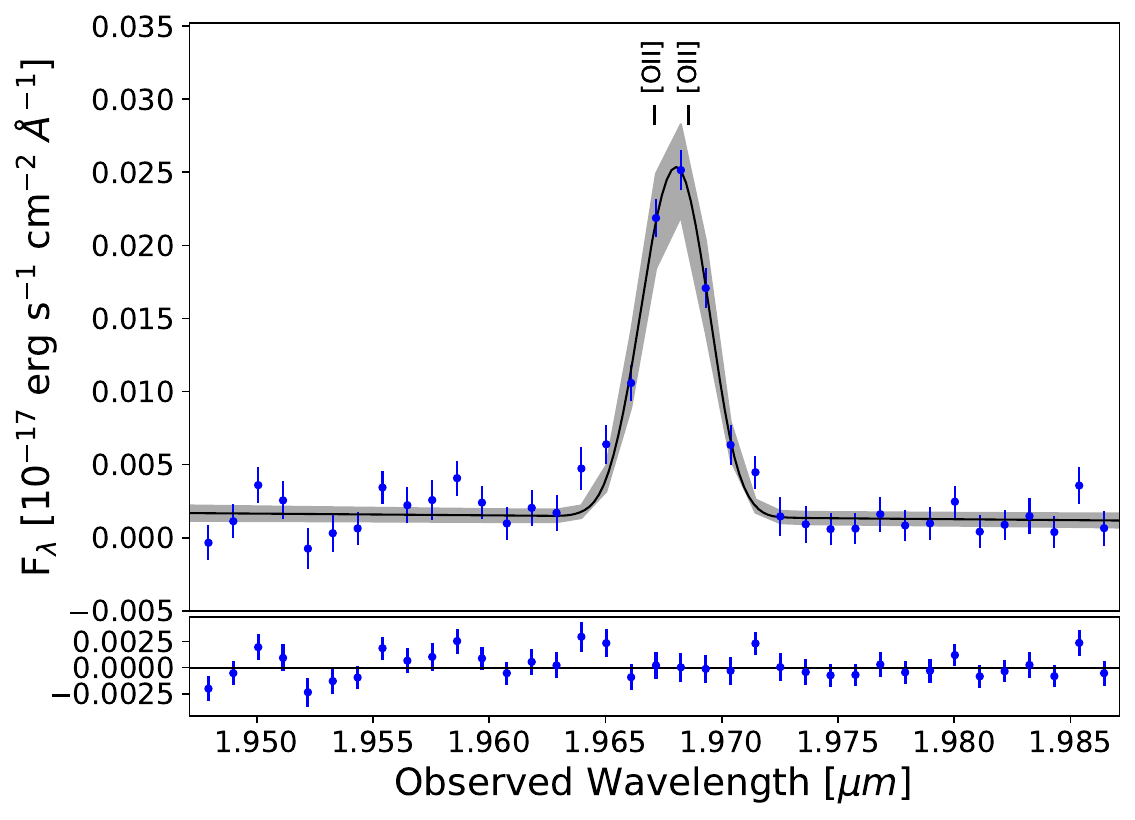}
 \hspace*{+.2cm}\includegraphics[width=0.5\textwidth]{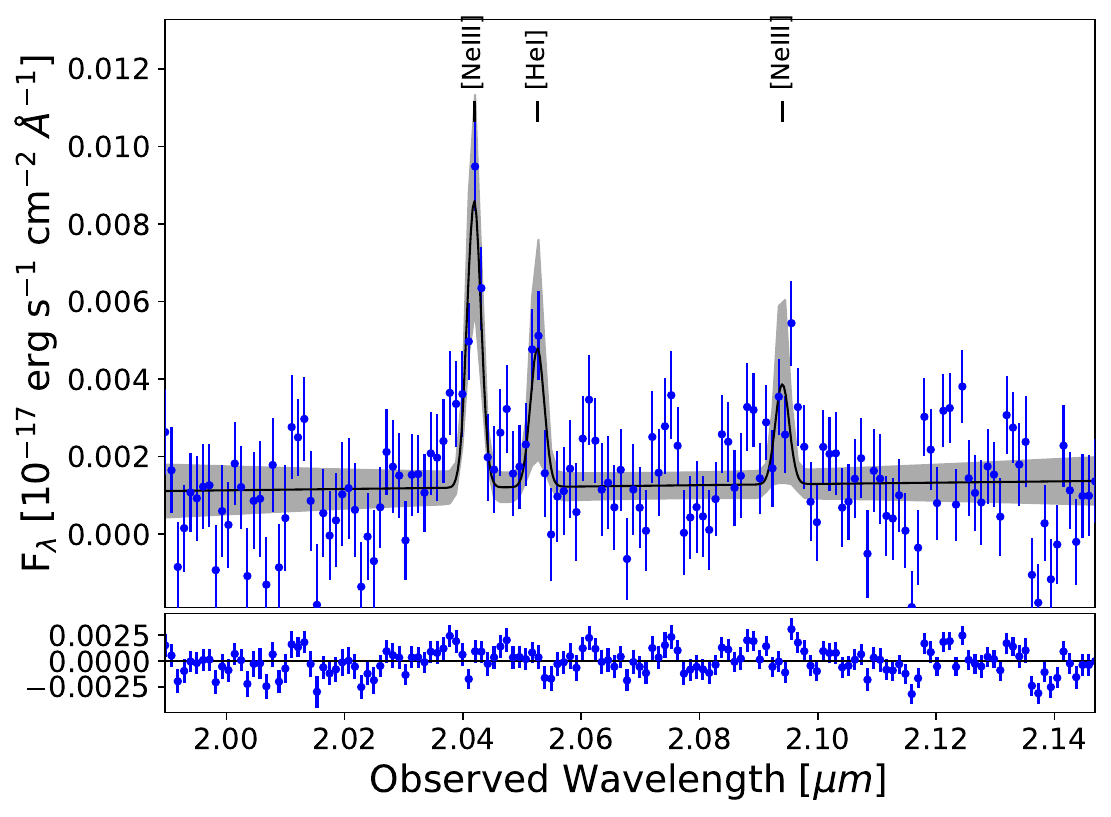}
 \hspace*{-.5cm}\includegraphics[width=0.5\textwidth]{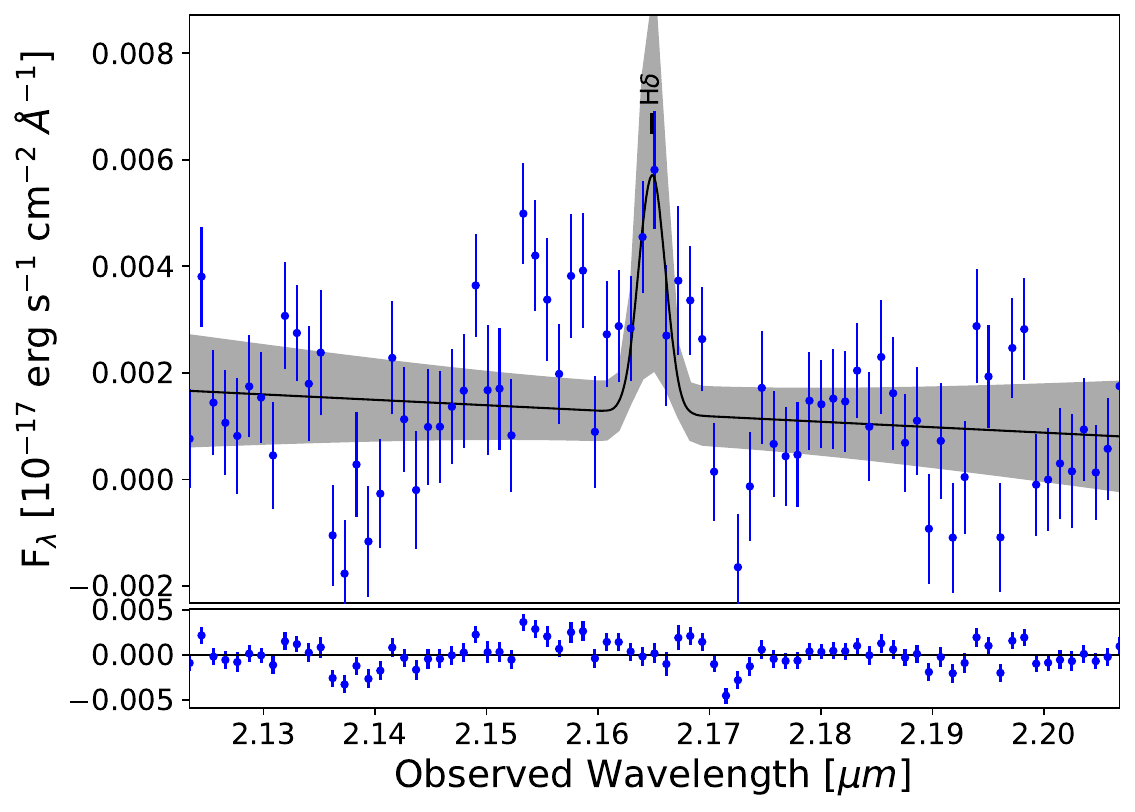}
 \hspace*{+.2cm}\includegraphics[width=0.5\textwidth]{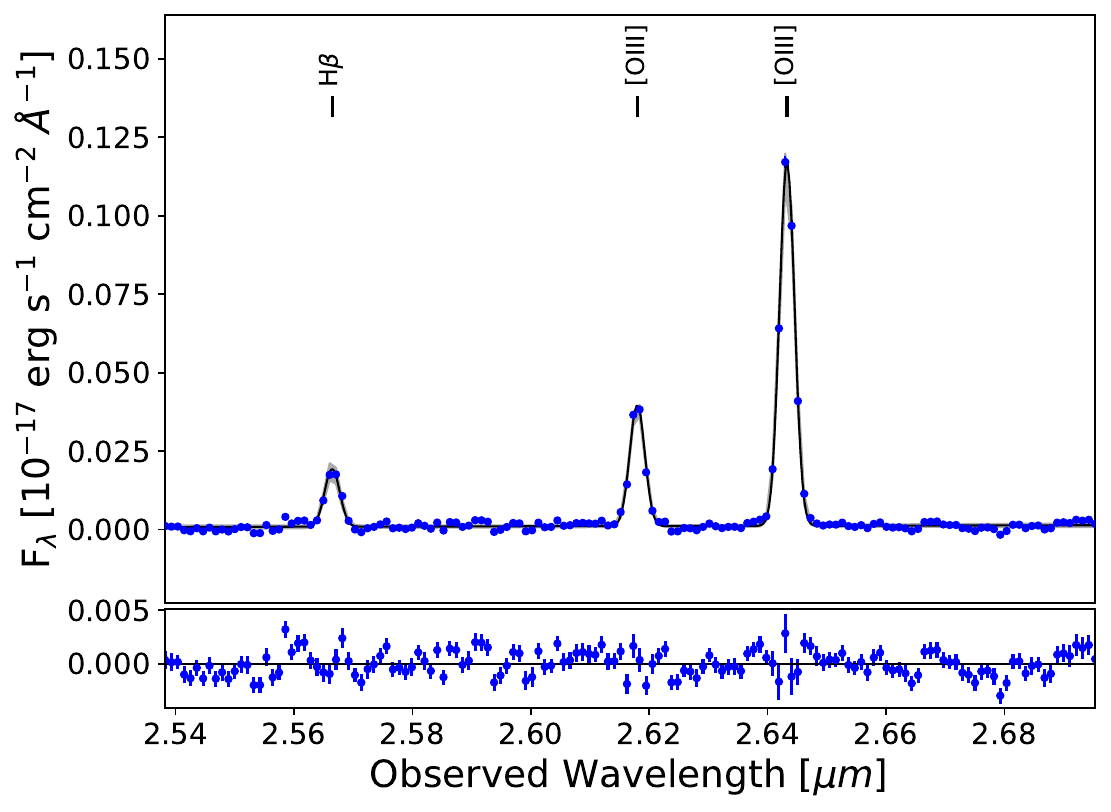}
 \hspace*{-.5cm}\includegraphics[width=0.5\textwidth]{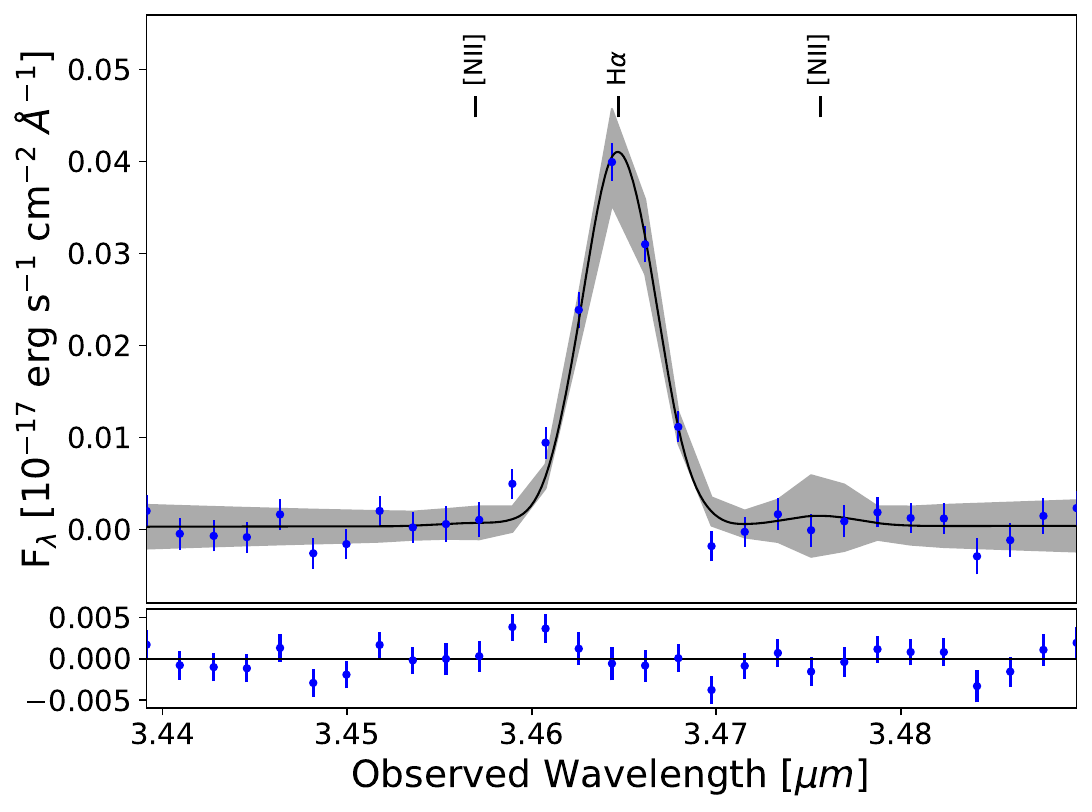}
 \hspace*{+.2cm}\includegraphics[width=0.5\textwidth]{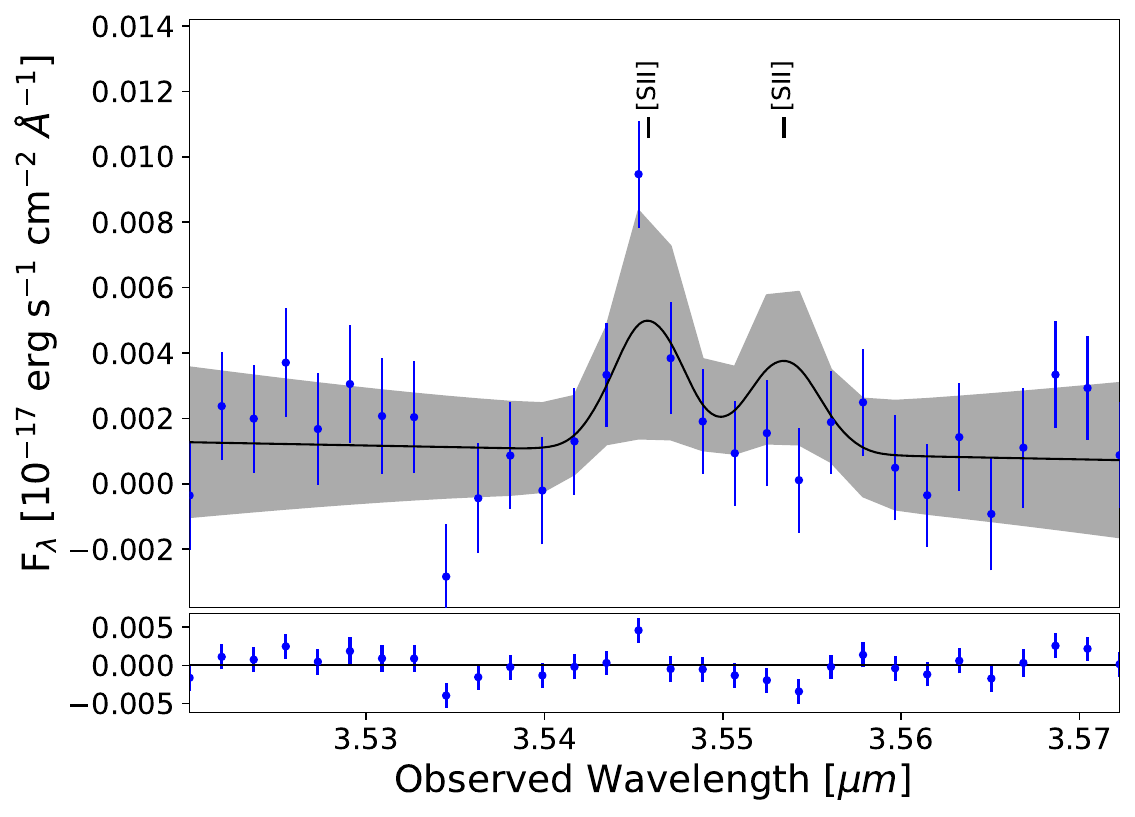} 
  \caption{As in Fig.~\ref{fig:OIIIA}, but for the remaining emission lines for the host of \protect\grbfive. The centroids of each fitted Gaussian is marked with a vertical dash and labelled. We note that we fit the \OII\ doublet with two individual Gaussians that blend into the line plotted in the first panel.
  }
 \label{apfig:line_fits}
\end{figure*}

In Fig.~\ref{apfig:line_fits} we show the best-fits to the remaining strong lines detected in our galaxy spectra described in Section~\ref{sec:line_fluxes}. The fluxes calculated from these fits are listed in Table~\ref{tab:line_fits}.

\section{Metallicity fits}\label{apsec:metallicites}

\begin{figure*}
 \centering
 \hspace*{-1.2cm}\includegraphics[width=0.35\textwidth]{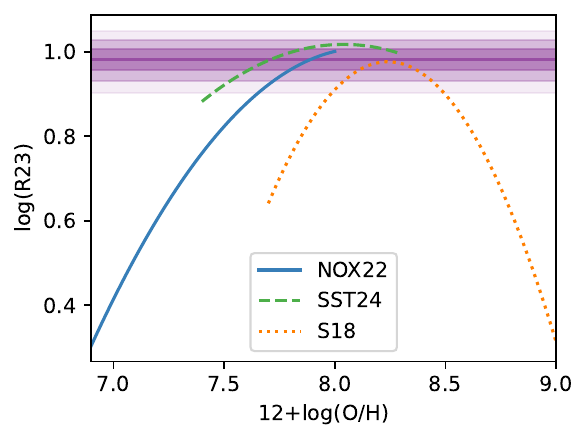}
 \hspace*{-.2cm}\includegraphics[width=0.35\textwidth]{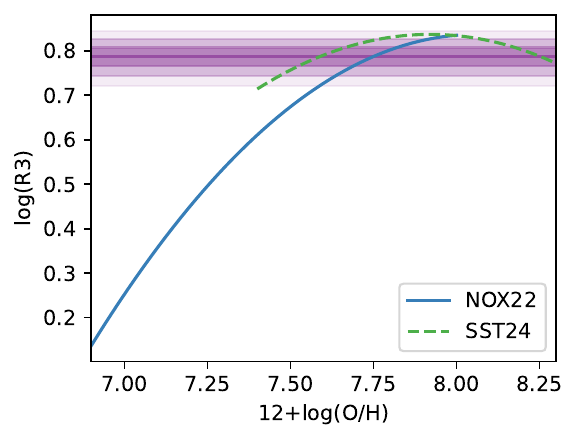}
 \hspace*{-.2cm}\includegraphics[width=0.35\textwidth]{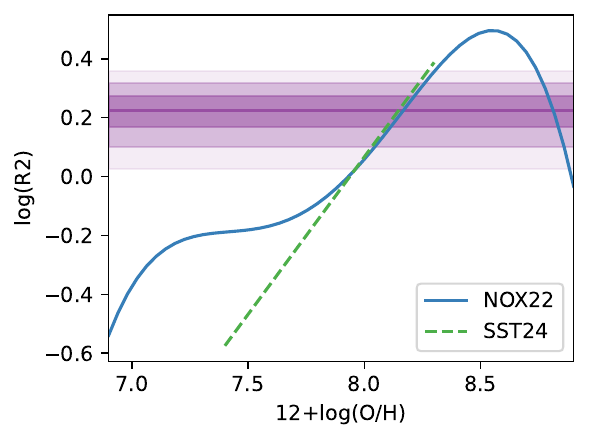}
 \hspace*{-1.2cm}\includegraphics[width=0.35\textwidth]{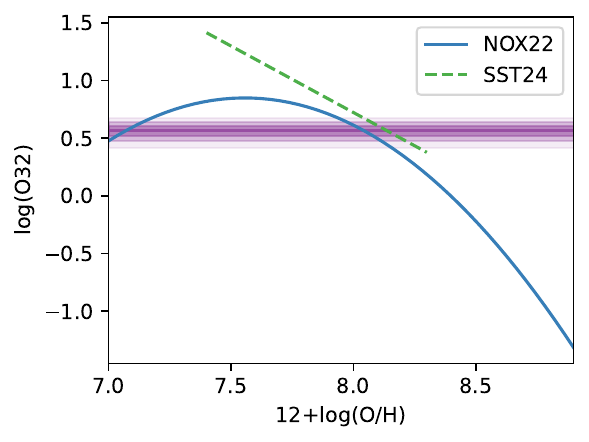}
 \hspace*{-.2cm}\includegraphics[width=0.35\textwidth]{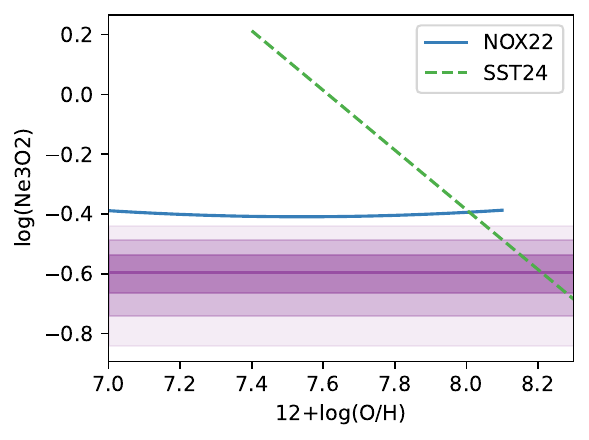}
 \hspace*{-.2cm}\includegraphics[width=0.35\textwidth]{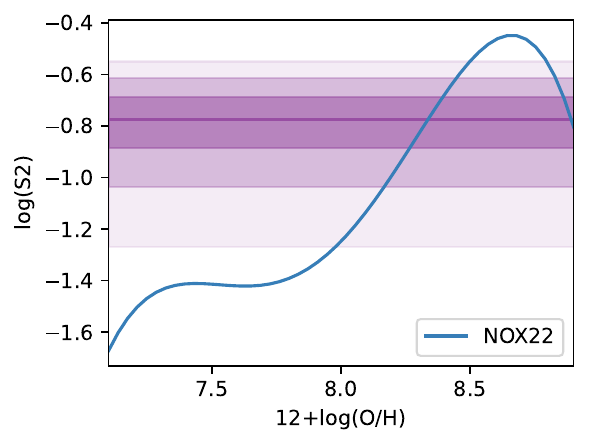} 
 \hspace*{-1.2cm}\includegraphics[width=0.35\textwidth]{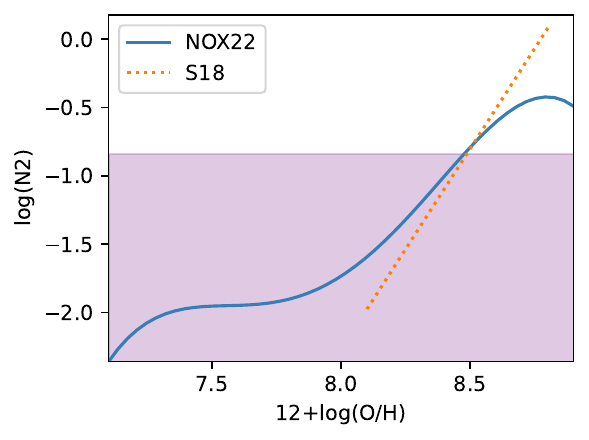} 
 \hspace*{-.0cm}\includegraphics[width=0.32\textwidth]{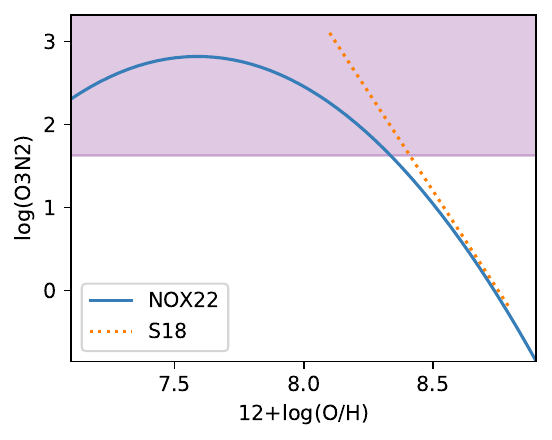}
 \hspace*{-.0cm}\includegraphics[width=0.35\textwidth]{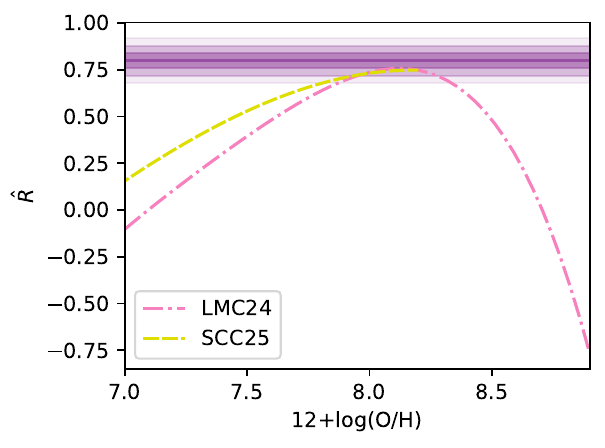} 
  \caption{Releations between strong lines ratios and metallicity. Different colours/linestyles represent different calibrations of the relations. The colours are the same as in Fig.~\ref{fig:z_comp}. The purple, horizontal line is te measured strong line ratio labelled on the y-axis. The shaded purple regions are 1$\sigma$, 2$\sigma$ and 3$\sigma$ errors on the strong line ratio. For $N_2$ and $O_3N_2$ only an upper or lower limit is determined, which is represented as the purple shaded area. The reported metallicities in Table~\ref{tab:metallicities} are the intersection points between the purple line and the various coloured lines. If the purple line lies outside the range of the calibrated relation, we give the maximum value of the curve as the corresponding value. For referece, the \Te-based metallicity we find is between 12 + log(O/H) = 7.80$\pm$0.19 and 7.96$\pm$0.21
  }
 \label{apfig:SL_fits}
\end{figure*}

In Fig.~\ref{apfig:SL_fits} we plot the relations between ratios of the strong emission lines considered in this paper and \Te-based metallicity. Each panel represents a different strong line ratio and each different coloured line (and style) represents a different calibration sample. See Sections~\ref{sec:SLdiags} of the main body of the paper for details on the line ratios and calibration samples used. The obtained metallicities are listed in Table~\ref{tab:metallicities}.

\bsp	
\label{lastpage}
\end{document}